\documentclass[twocolumn]{aastex63}
\usepackage{graphicx,amsmath,amsthm}

\shorttitle{Evolution of Rotation and Magnetic Activity}
\shortauthors{Metcalfe et al.}

\journalinfo{The Astrophysical Journal, accepted}

\begin{document}

\title{\bf The Evolution of Rotation and Magnetic Activity in 94\,Aqr\,Aa from Asteroseismology with TESS}

\author[0000-0003-4034-0416]{Travis S.~Metcalfe}
\affiliation{Space Science Institute, 4765 Walnut St., Suite B, Boulder, CO 80301, USA}
\affiliation{White Dwarf Research Corporation, 3265 Foundry Pl., Unit 101, Boulder, CO 80301, USA}
\affiliation{Max-Planck-Institut f\"ur Sonnensystemforschung, Justus-von-Liebig-Weg 3, 37077, G\"ottingen, Germany}

\author[0000-0002-4284-8638]{Jennifer L.~van~Saders}
\affiliation{Institute for Astronomy, University of Hawai`i, 2680 Woodlawn Drive, Honolulu, HI 96822, USA}

\author[0000-0002-6163-3472]{Sarbani Basu}
\affiliation{Department of Astronomy, Yale University, PO Box 208101, New Haven, CT 06520-8101, USA}

\author[0000-0002-1988-143X]{Derek Buzasi}
\affiliation{Department of Chemistry and Physics, Florida Gulf Coast University, 10501 FGCU Blvd S, Fort Myers, FL 33965}

\author[0000-0002-5714-8618]{William J.~Chaplin}
\affiliation{School of Physics \& Astronomy, University of Birmingham, Edgbaston, Birmingham B15 2TT, UK}
\affiliation{Stellar Astrophysics Centre, Aarhus University, Ny Munkegade 120, DK-8000 Aarhus C, Denmark}

\author[0000-0002-4996-0753]{Ricky Egeland}
\affiliation{High Altitude Observatory, National Center for Atmospheric Research, P.O. Box 3000, Boulder, CO 80307-3000, USA}

\author[0000-0002-8854-3776]{Rafael A.~Garcia}
\affiliation{IRFU, CEA, Universit\'e Paris-Saclay, F-91191 Gif-sur-Yvette, France}
\affiliation{AIM, CEA, CNRS, Universit\'e Paris-Saclay, Universit\'e Paris Diderot, Sorbonne Paris Cit\'e, F-91191 Gif-sur-Yvette, France}

\author[0000-0001-8330-5464]{Patrick Gaulme}
\affiliation{Max-Planck-Institut f\"ur Sonnensystemforschung, Justus-von-Liebig-Weg 3, 37077, G\"ottingen, Germany}

\author[0000-0001-8832-4488]{Daniel Huber}
\affiliation{Institute for Astronomy, University of Hawai`i, 2680 Woodlawn Drive, Honolulu, HI 96822, USA}

\author[0000-0002-1299-1994]{Timo Reinhold}
\affiliation{Max-Planck-Institut f\"ur Sonnensystemforschung, Justus-von-Liebig-Weg 3, 37077, G\"ottingen, Germany}

\author[0000-0001-9932-9559]{Hannah Schunker}
\affiliation{Max-Planck-Institut f\"ur Sonnensystemforschung, Justus-von-Liebig-Weg 3, 37077, G\"ottingen, Germany}
\affiliation{School of Mathematical and Physical Sciences, University of Newcastle, Callaghan, New South Wales, Australia}

\author[0000-0002-3481-9052]{Keivan G.~Stassun} 
\affiliation{Vanderbilt University, Department of Physics \& Astronomy, 6301 Stevenson Center Lane, Nashville, TN 37235, USA}

\author[0000-0002-1790-1951]{Thierry Appourchaux}
\affiliation{Institut d'Astrophysique Spatiale, UMR8617, B\^atiment 121, 91045 Orsay Cedex, France}

\author[0000-0002-4773-1017]{Warrick H.~Ball}
\affiliation{School of Physics \& Astronomy, University of Birmingham, Edgbaston, Birmingham B15 2TT, UK}
\affiliation{Stellar Astrophysics Centre, Aarhus University, Ny Munkegade 120, DK-8000 Aarhus C, Denmark}

\author[0000-0001-5222-4661]{Timothy R.~Bedding}
\affiliation{Sydney Institute for Astronomy, School of Physics, University of Sydney 2006, Australia}
\affiliation{Stellar Astrophysics Centre, Aarhus University, Ny Munkegade 120, DK-8000 Aarhus C, Denmark}

\author[0000-0003-4317-5460]{S\'ebastien Deheuvels}
\affiliation{IRAP, Université de Toulouse, CNRS, CNES, UPS, Toulouse, France}

\author[0000-0002-1241-5508]{Luc\'{\i}a Gonz\'alez-Cuesta}
\affiliation{Instituto de Astrof\'{\i}sica de Canarias, La Laguna, Tenerife, Spain}
\affiliation{Dpto. de Astrof\'{\i}sica, Universidad de La Laguna, La Laguna, Tenerife, Spain}

\author[0000-0001-8725-4502]{Rasmus Handberg}
\affiliation{Stellar Astrophysics Centre, Aarhus University, Ny Munkegade 120, DK-8000 Aarhus C, Denmark}

\author[0000-0003-4480-6690]{Antonio Jim\'enez}
\affiliation{Instituto de Astrof\'{\i}sica de Canarias, La Laguna, Tenerife, Spain}
\affiliation{Dpto. de Astrof\'{\i}sica, Universidad de La Laguna, La Laguna, Tenerife, Spain}

\author[0000-0002-9037-0018]{Hans Kjeldsen}
\affiliation{Stellar Astrophysics Centre, Aarhus University, Ny Munkegade 120, DK-8000 Aarhus C, Denmark}
\affiliation{Institute of Theoretical Physics and Astronomy, Vilnius University, Sauletekio av. 3, 10257 Vilnius, Lithuania}

\author[0000-0001-6396-2563]{Tanda Li}
\affiliation{Sydney Institute for Astronomy, School of Physics, University of Sydney 2006, Australia}
\affiliation{Stellar Astrophysics Centre, Aarhus University, Ny Munkegade 120, DK-8000 Aarhus C, Denmark}
\affiliation{Key Laboratory of Solar Activity, National Astronomical Observatories, Chinese Academy of Science, Beijing 100012, China}

\author[0000-0001-9214-5642]{Mikkel N.~Lund}
\affiliation{Stellar Astrophysics Centre, Aarhus University, Ny Munkegade 120, DK-8000 Aarhus C, Denmark}

\author[0000-0002-0129-0316]{Savita Mathur}
\affiliation{Instituto de Astrof\'{\i}sica de Canarias, La Laguna, Tenerife, Spain}
\affiliation{Dpto. de Astrof\'{\i}sica, Universidad de La Laguna, La Laguna, Tenerife, Spain}

\author[0000-0002-7547-1208]{Benoit Mosser}
\affiliation{LESIA, Observatoire de Paris, Universit\'e PSL, CNRS, Sorbonne Universit\'e, Universit\'e de Paris, 92195 Meudon, France}

\author[0000-0001-9169-2599]{Martin B.~Nielsen}
\affiliation{School of Physics \& Astronomy, University of Birmingham, Edgbaston, Birmingham B15 2TT, UK}
\affiliation{Stellar Astrophysics Centre, Aarhus University, Ny Munkegade 120, DK-8000 Aarhus C, Denmark}
\affiliation{Center for Space Science, NYUAD Institute, New York University Abu Dhabi, PO Box 129188, Abu Dhabi, United Arab Emirates}

\author[0000-0002-6957-6871]{Anthony Noll}
\affiliation{IRAP, Université de Toulouse, CNRS, CNES, UPS, Toulouse, France}

\author[0000-0002-9424-2339]{Zeynep \c{C}elik Orhan}
\affiliation{Department of Astronomy and Space Sciences, Science Faculty, Ege University, 35100 Bornova, \.Izmir, Turkey}

\author[0000-0001-5759-7790]{Sibel \"Ortel}
\affiliation{Department of Astronomy and Space Sciences, Science Faculty, Ege University, 35100 Bornova, \.Izmir, Turkey}

\author[0000-0001-7195-6542]{\^Angela R.~G.~Santos}
\affiliation{Space Science Institute, 4765 Walnut St., Suite B, Boulder, CO 80301, USA}

\author[0000-0002-7772-7641]{Mutlu Yildiz}
\affiliation{Department of Astronomy and Space Sciences, Science Faculty, Ege University, 35100 Bornova, \.Izmir, Turkey}

\author{Sallie Baliunas}
\affiliation{Harvard-Smithsonian Center for Astrophysics, Cambridge, MA 02138, USA}

\author{Willie Soon}
\affiliation{Harvard-Smithsonian Center for Astrophysics, Cambridge, MA 02138, USA}

\begin{abstract}

Most previous efforts to calibrate how rotation and magnetic activity 
depend on stellar age and mass have relied on observations of clusters, 
where isochrones from stellar evolution models are used to determine the 
properties of the ensemble. Asteroseismology employs similar models to 
measure the properties of an individual star by matching its normal modes 
of oscillation, yielding the stellar age and mass with high precision. We 
use 27 days of photometry from the Transiting Exoplanet Survey 
Satellite to characterize solar-like oscillations in the G8 
subgiant of the 94~Aqr triple system. The resulting stellar properties, 
when combined with a reanalysis of 35 yr of activity measurements from 
the Mount Wilson HK project, allow us to probe the evolution of rotation 
and magnetic activity in the system. The asteroseismic age of the 
subgiant agrees with a stellar isochrone fit, but the rotation period is 
much shorter than expected from standard models of angular momentum 
evolution. We conclude that weakened magnetic braking may be needed to 
reproduce the stellar properties, and that evolved subgiants in the 
hydrogen shell-burning phase can reinvigorate large-scale dynamo action 
and briefly sustain magnetic activity cycles before ascending the red 
giant branch.

\end{abstract}


\NewPageAfterKeywords

\section{Introduction}\label{sec1}

Studies of long-term magnetic variability in solar-type stars rely on 
measurements of chromospheric activity obtained over many decades. 
Fortunately, the collection of such observations started in the late 
1960s from the Mount Wilson Observatory \citep{Wilson1978} and continued 
for more than 35 years. A similar program at Lowell Observatory 
\citep{Hall2007} began in the early 1990s and is still ongoing, with the 
composite time-series for some stars now approaching half a century 
\citep{Egeland2017}. With sufficiently frequent sampling during each 
observing season, the modulation from individual active regions can 
reveal the stellar rotation period \citep{Baliunas1983}, while changes 
between seasons can constrain latitudinal differential rotation from the 
slow migration of active regions through the magnetic cycle 
\citep{Donahue1996}. Such long-term data sets have provided high-quality 
snapshots of magnetic variability in dozens of solar-type stars 
\citep{BohmVitense2007, Brandenburg2017}, but the evolutionary thread 
that connects them is difficult to establish due to uncertainties in the 
basic stellar properties such as mass and age \citep{Metcalfe2017}.

Asteroseismology with the {\it Transiting Exoplanet Survey Satellite} 
\citep[TESS,][]{Ricker2014} is poised to revolutionize our understanding 
of the evolution of magnetic variability in solar-type stars. It provides 
nearly uninterrupted time-series photometry with a 2-minute cadence 
spanning at least 27 days, which is sufficient to detect solar-like 
oscillations in many F- and G-type dwarfs and subgiants down to V$\sim$7 
\citep{Schofield2019}. The detection of global oscillation properties 
such as the frequency of maximum oscillation power ($\nu_{\rm max}$) and 
the mean frequency spacing between consecutive radial overtones 
($\Delta\nu$), when combined with spectroscopic properties such as the 
effective temperature and metallicity ($T_{\rm eff}$, [Fe/H]), can 
typically determine the stellar mass with an uncertainty of 6\% and the 
stellar age within about 20\% \citep{Serenelli2017}. The identification 
of individual oscillation frequencies can cut these uncertainties in half 
\citep{Creevey2017}, and even larger improvements in the age precision 
are possible for subgiants that exhibit mixed-modes, which couple 
gravity-driven $g$ mode oscillations in the stellar core with 
pressure-driven $p$ mode oscillations in the envelope 
\citep{Deheuvels2011, Li2019}.

In this paper, we demonstrate the power of combining ground-based 
magnetic variability data with asteroseismic measurements of basic 
stellar properties from TESS. Our initial application is to the 94~Aqr 
triple system (HD\,219834), which includes a blended primary consisting 
of a G8 subgiant (Aa) and a K3 dwarf (Ab) in a 6.3~year orbit, and a 
resolved secondary K2 dwarf (B) separated by 13 arcseconds 
\citep{Fuhrmann2008}. In Section~\ref{sec2} we provide an overview of the 
observations, and in Section~\ref{sec3} we reanalyze the archive of 
chromospheric activity measurements from the Mount Wilson survey 
\citep{Baliunas1995} for both the blended primary (A) and the resolved 
secondary (B) to determine the activity cycle and rotation periods. All 
three components are blended in the TESS observations, but the subgiant 
produces the only detectable asteroseismic signal because the K dwarfs 
oscillate with a much lower amplitude and higher frequency. In 
Section~\ref{sec4} we analyze and model the subgiant oscillations to 
determine the basic stellar properties from asteroseismology, and in 
Section~\ref{sec5} we establish the accuracy of these results with 
independent estimates of the subgiant radius, mass and age. In 
Section~\ref{sec6} we combine the rotation period from Mount Wilson data 
with the stellar mass and age from TESS to model the angular momentum 
evolution of the subgiant, demonstrating that weakened magnetic braking 
\citep{vanSaders2016} may be needed to explain the current rotation 
period. Finally, in Section~\ref{sec7} we summarize and discuss our 
results, including the idea that evolved subgiants can sustain a 
``born-again'' dynamo before ascending the red giant branch.

\section{Observations}\label{sec2}

\subsection{Mount Wilson HK data}\label{sec2.1}

We use observations from the Mount Wilson Observatory (MWO) HK project 
\citep{Wilson1978, Vaughan1978, Baliunas1995} to study the long- and 
short-term variability of magnetic activity in 94~Aqr~A and B. The HK 
Project HKP-1 (1966--1978) and HKP-2 (1978--2003) spectrophotometers 
obtained counts through 1~\AA\ triangular bandpasses centered on the 
Ca~{\sc ii} H \& K (hereafter HK) line cores at 3968.470~\AA\ and 
3933.664~\AA, respectively, as well as two 20~\AA\ pseudo-continuum 
bands, $R$ centered at 4001.067~\AA, and $V$ centered at 3901.068~\AA\ 
\citep{Vaughan1978}. Emission in the HK line cores has long been known to 
be a signature of surface magnetic flux \citep[see][for a 
review]{Linsky1970}, and the disk-integrated HK emission from the Sun 
reveals the solar cycle \citep[e.g.][]{White1981, Egeland2017b}. The 
ratio of core to pseudo-continuum counts, $S = \alpha_{\rm MWO} (N_H + 
N_K)/(N_R + N_V)$, where $\alpha_{\rm MWO}$ is a calibration factor, 
defines the now-standard $S$-index of magnetic activity. The pair 
94~Aqr~Aa and Ab were not resolved in the MWO observations, so the 
94~Aqr~A time series represents the sum of surface fluxes from these 
components. However, the G8 subgiant contributes approximately 97\% of 
the flux in the relevant bandpasses (see Section~\ref{sec5.1}).

\subsection{TESS photometry}\label{sec2.2}

TESS observed 94~Aqr in 2-minute cadence for 27 days during Sector 2 of 
Cycle 1 (2018 Aug 22--2018 Sep 20). We used the target pixel files 
produced by the TESS Science Processing Operations Center 
\citep{Jenkins2016} to extract light curves. A preliminary detection of 
solar-like oscillations was made with a light curve produced using simple 
aperture photometry, selecting all pixels with flux above three times the 
median absolute deviation of a median stacked image over the full 
observing sector. The final light curve was produced using the photometry 
pipeline\footnote{\url{https://tasoc.dk/code/}} (Handberg et al., in 
prep.) maintained by the TESS Asteroseismic Science Operations Center 
\citep[TASOC,][]{Lund2017}, which is based on software originally 
developed to generate light curves from data collected by the {\it K2} 
mission \citep{Lund2015}.

 \begin{figure}
 \centering\includegraphics[width=\columnwidth]{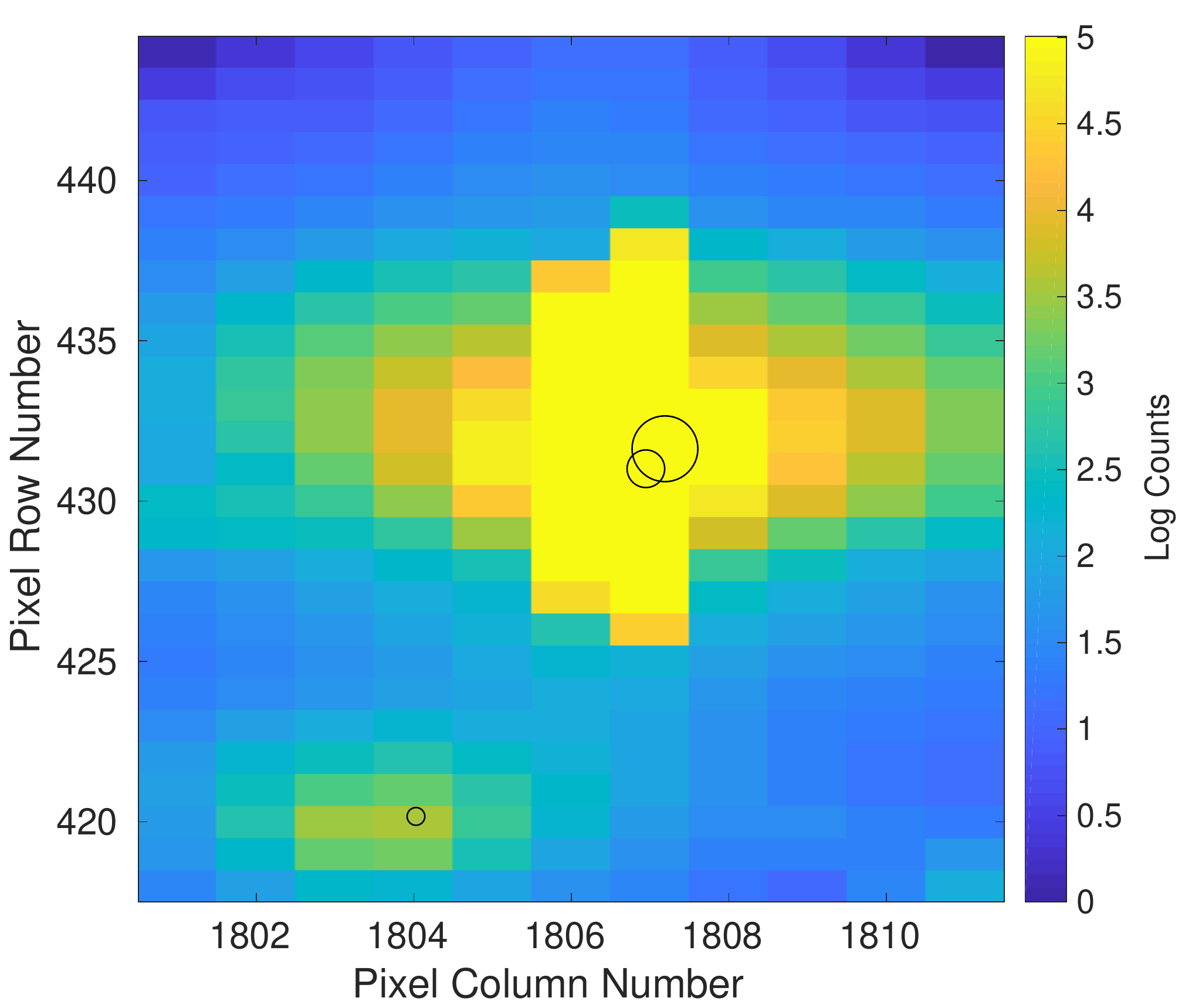} 
 \caption{The mean TESS postage stamp ($3\farcm9\times9\farcm1$) for 
94~Aqr, averaged over the complete time series. Counts are shown on a 
logarithmic scale to allow visibility of the full range, and overlying 
circles indicate stellar locations from {\it Gaia} DR2, with a limiting G 
magnitude of 14. Blending of the A and B components near image center is 
apparent.\label{fig1}}
 \end{figure} 
 
Figure \ref{fig1} shows the postage stamp for 94~Aqr. The large TESS 
pixels mean that the components of the 94~Aqr system are separated by 
less than one pixel on the detector. However, we made an effort to 
extract separate light curves for the two components. We built custom 
aperture masks around well-separated portions of the combined stellar 
image, conducted photometric extractions using those masks, and detrended 
the resulting light curves against spacecraft pointing data using a 
second-order two-dimensional polynomial fit, which has worked well for 
{\it Kepler} data in the past \citep{Buzasi2015}. Our goal was to 
construct aperture masks which were dominated by the wings of the images 
of the two stars, which might then allow us to separate the two stellar 
contributions. Despite our detrending efforts, light curves resulting 
from such aperture masks tended to be dominated by photometric jitter 
resulting from spacecraft motion, and we were unable to unequivocally 
separate the target light curves using this approach.

\subsection{Derived luminosity}\label{sec2.3}

We derived an updated luminosity for 94~Aqr~Aa from speckle observations 
of the close ($0\farcs15$) binary A component and the {\it Gaia} DR2 
parallax of the resolved B component (see Section~\ref{sec5.2}). The 
total V magnitude of the A component is 5.18\,$\pm$\,0.01 
\citep{Fabricius2002}, while the magnitude difference between Aa and Ab 
from speckle imaging is 3.1 \citep{Tokovinin2015}. The bolometric 
correction ($-$0.04) was deduced from an extrapolation of \citet[][their 
Fig.~26]{VandenBerg2003}, adopting $T_{\rm eff}=5461\pm40$~K from 
\cite{Gray2006}\footnote{Note that the quoted uncertainty on $T_{\rm 
eff}$ does not account for systematics between different methods and the 
fundamental $T_{\rm eff}$ scale set by the accuracy of interferometric 
angular diameters, which can be $\ga$2\% \citep{Casagrande2014, 
White2018}.}, the parallax of 94~Aqr~B adjusted for a small systematic 
offset \citep[e.g.,][]{StassunTorres2018, Zinn2018}, and interstellar 
extinction $A_V = 0.02^{+0.07}_{-0.02}$ (see Section~\ref{sec5.1}). The 
luminosity deduced from these parameters is $L_{\rm 
Aa}=3.31_{-0.07}^{+0.22}\ L_\odot$.

\section{Activity Cycles and Rotation}\label{sec3}

The resolved components of the 94~Aqr system (A and B) were observed by 
the Mount Wilson program during the years 1967--2003. The $S$-index was 
measured during annual observing seasons, each covering time spans 
between two and six months. From these data sets, we used several methods 
to estimate the activity cycle period and the rotation period for each 
star.

\begin{figure} 
 \centering\includegraphics[width=\columnwidth]{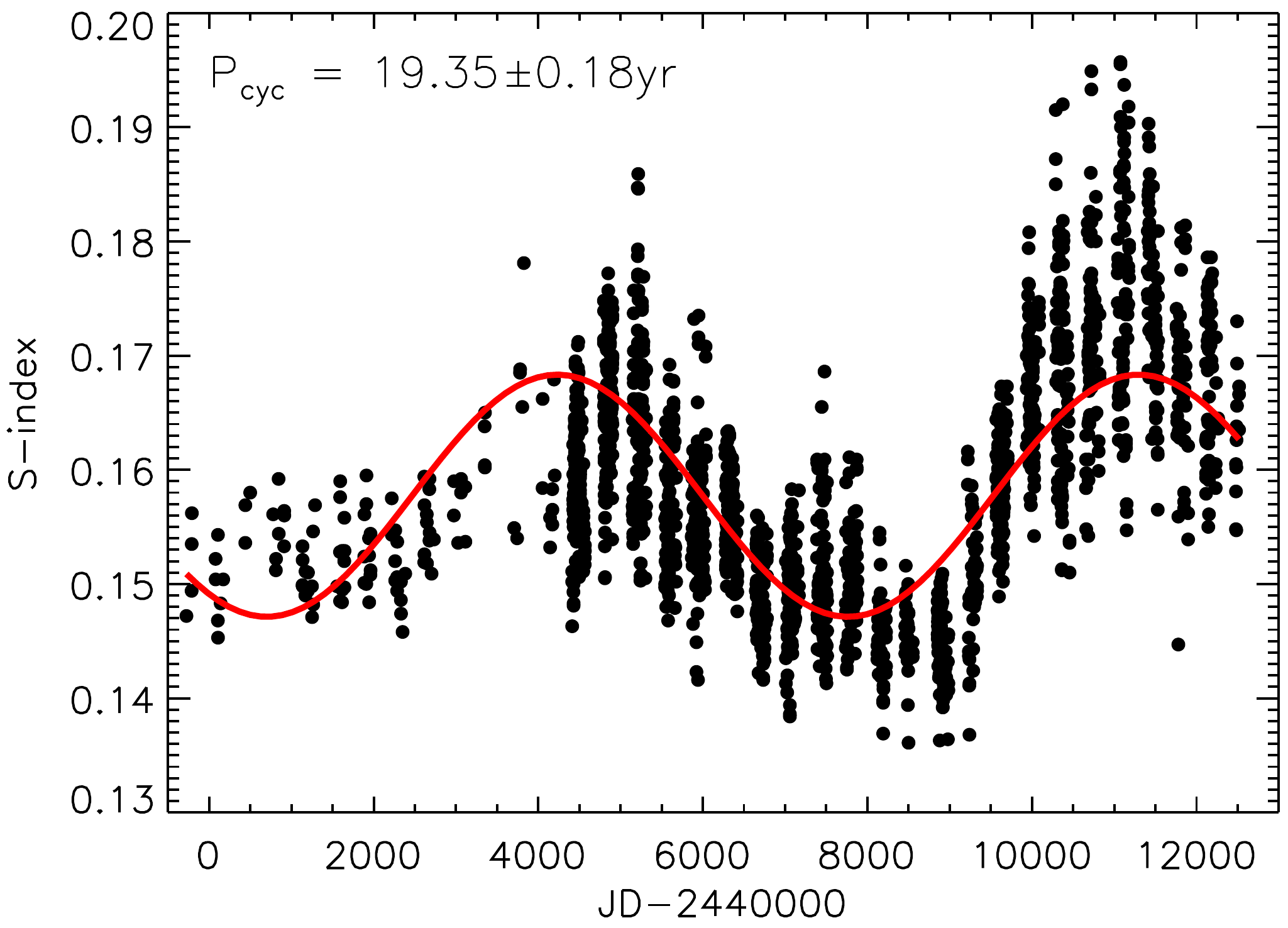}
 \centering\includegraphics[width=\columnwidth]{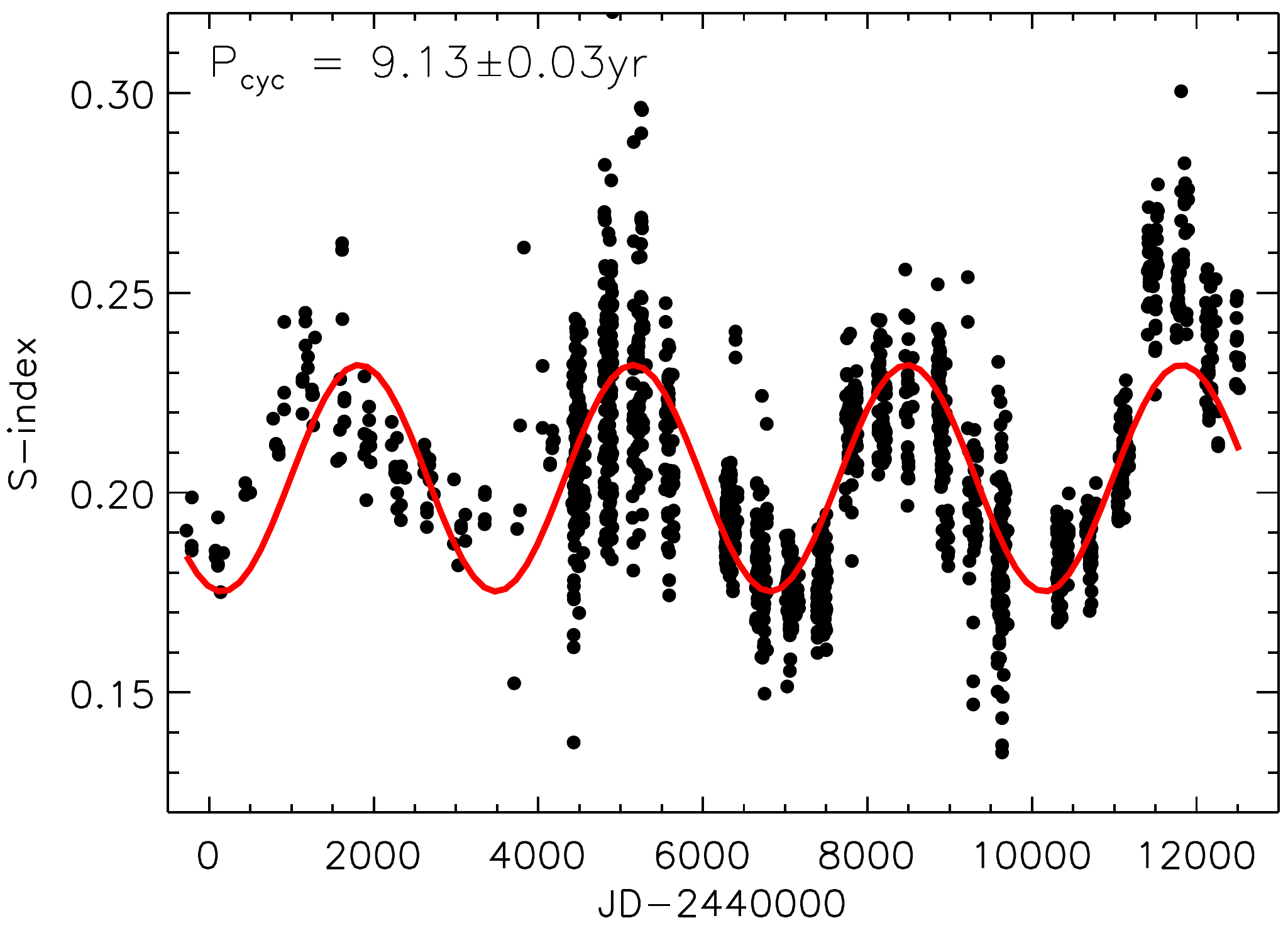} 
 \caption{Long-term variability of 94~Aqr~A (top) and 94~Aqr~B (bottom) 
from the Mount Wilson $S$-index data. The reported activity cycle period 
and uncertainty come from fitting a sinusoid (red curves) to each time 
series.\label{fig2}}
 \end{figure}

Each $S$-index time series shows long-term variability due to magnetic 
activity cycles. We computed the Lomb-Scargle periodogram of the full 
$S$-index time series for each component to search for the activity cycle 
period. We fit a sinusoid to the data using the highest periodogram peak 
as the initial period guess. The peak heights at the derived periods were 
used to estimate the period uncertainties. Following 
\cite{Montgomery1999}, we estimated the uncertainty of the peak with 
frequency $f$ as $\sigma(f)=\sqrt{6/N}\cdot\sigma(m)/(\pi T a)$, where 
$N$ is the number of data points, $T$ is the time baseline of the 
observations, and $a$ is the amplitude of the sinusoid. We computed the 
root-mean-square deviation $\sigma(m)=\sqrt{\sigma^2\cdot(1-h_{\rm 
peak})}$, where $\sigma^2$ is the variance of the zero-mean time series, 
and $h_{\rm peak}$ is the normalized peak height at the derived period. 
We obtained activity cycle periods of $19.35\pm0.18$~yr and 
$9.13\pm0.03$~yr for 94~Aqr~A and B, respectively (see 
Figure~\ref{fig2}). We validated this analysis with two other methods, 
the autocorrelation function and a time-period analysis using a Morlet 
wavelet \citep{Mathur2010, Garcia2014, Buzasi2016}. These analyses gave 
similar results.

\begin{figure}
 \centering\includegraphics[width=\columnwidth]{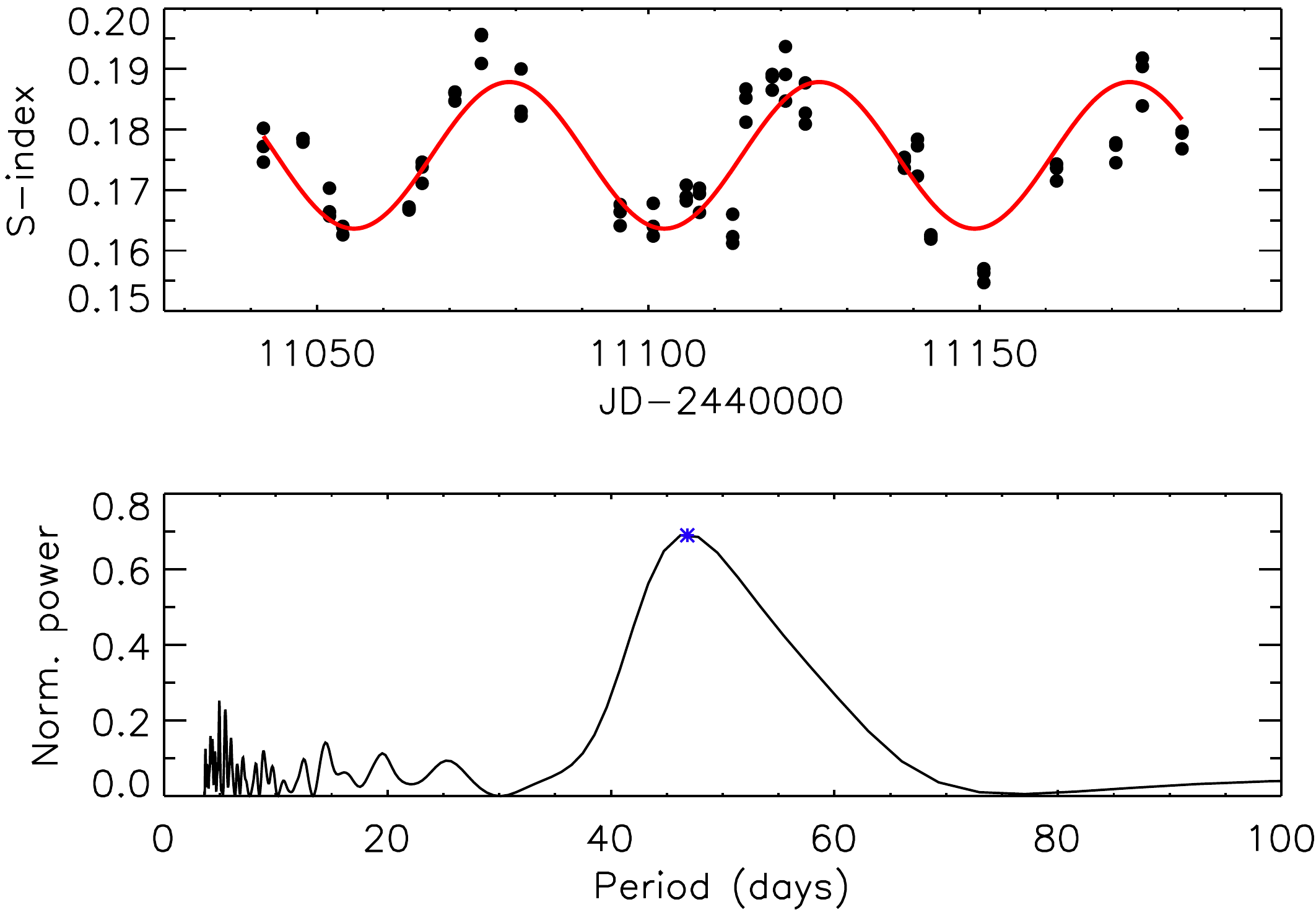} 
 \caption{Time series Ca~HK measurements of 94~Aqr~A from a representative
 observing season near cycle maximum in 1998 (top). The corresponding
 periodogram (bottom) shows a clear detection of rotation (blue point)
 with a period similar to the median value reported in the text. The
 detected period is shown with a red line in the top panel.\label{fig3}}
 \end{figure}

The number of data points per observing season at Mount Wilson increased 
around 1980, revealing large variability which we attribute to active 
regions rotating in and out of view. From the seasonal data, 
\cite{Baliunas1996} measured rotation periods of 42~days and 43~days for 
the A and B components, while \cite{Olspert2018} found $43.4\pm1.9$~days 
and $34.8\pm0.9$~days, respectively. We performed an independent analysis 
of the Mount Wilson data sets, with three teams analyzing them seasonally 
and four teams analyzing the complete time series globally. The teams 
applied a range of different analysis techniques to both stars in the 
system, including application of the discrete Fourier transform (DFT), 
performing sinusoid fits to individual seasons and studying a histogram 
of the results, applying autocorrelation functions, and time-period 
wavelet analysis. In the case of 94~Aqr~A, there was excellent agreement 
amongst the different methods, and we report a median rotation period of 
$P_{\rm rot}=46.9\pm1.9$~days (see Figure~\ref{fig3}). For 94~Aqr~B,
the teams could not reach a consensus on the statistical significance of
any potential detections of rotation from the seasonal data sets.

For the resolved B component, the activity cycle period can be attributed 
unambiguously to the K2 dwarf. For the blended A component, the G8 
subgiant (Aa) contributes approximately 97\% of the flux in the relevant 
bandpasses (see Section~\ref{sec5.1}). Consequently, the 3\% of the flux 
contributed by the K3 dwarf (Ab) would need to vary by roughly an order 
of magnitude to explain the observed $S$-index variation of up to 30\% on 
both rotational and activity cycle timescales. Such modulations would be 
unprecedented \citep[e.g., see][]{Soon1994}, so we identify the G8 
subgiant as the source of the $S$-index variability. Note that
comparable rotation periods have been observed in other G-type subgiants,
including HD\,182572 \citep[41~days,][]{Baliunas1996}, and KIC\,8524425
\citep[$42\pm3$~days,][]{Garcia2014}. The implications of 
the observed activity cycle, and the possibility of a ``born-again'' 
dynamo in evolved subgiants, are discussed in Section~\ref{sec7}.

\section{Asteroseismology of 94 A\lowercase{qr} A\lowercase{a}}\label{sec4}

\subsection{Extracting the oscillation parameters}\label{sec4.1}

\begin{figure} 
 \centering\includegraphics[angle=0,width=0.96\columnwidth,trim=21 2 0 8]{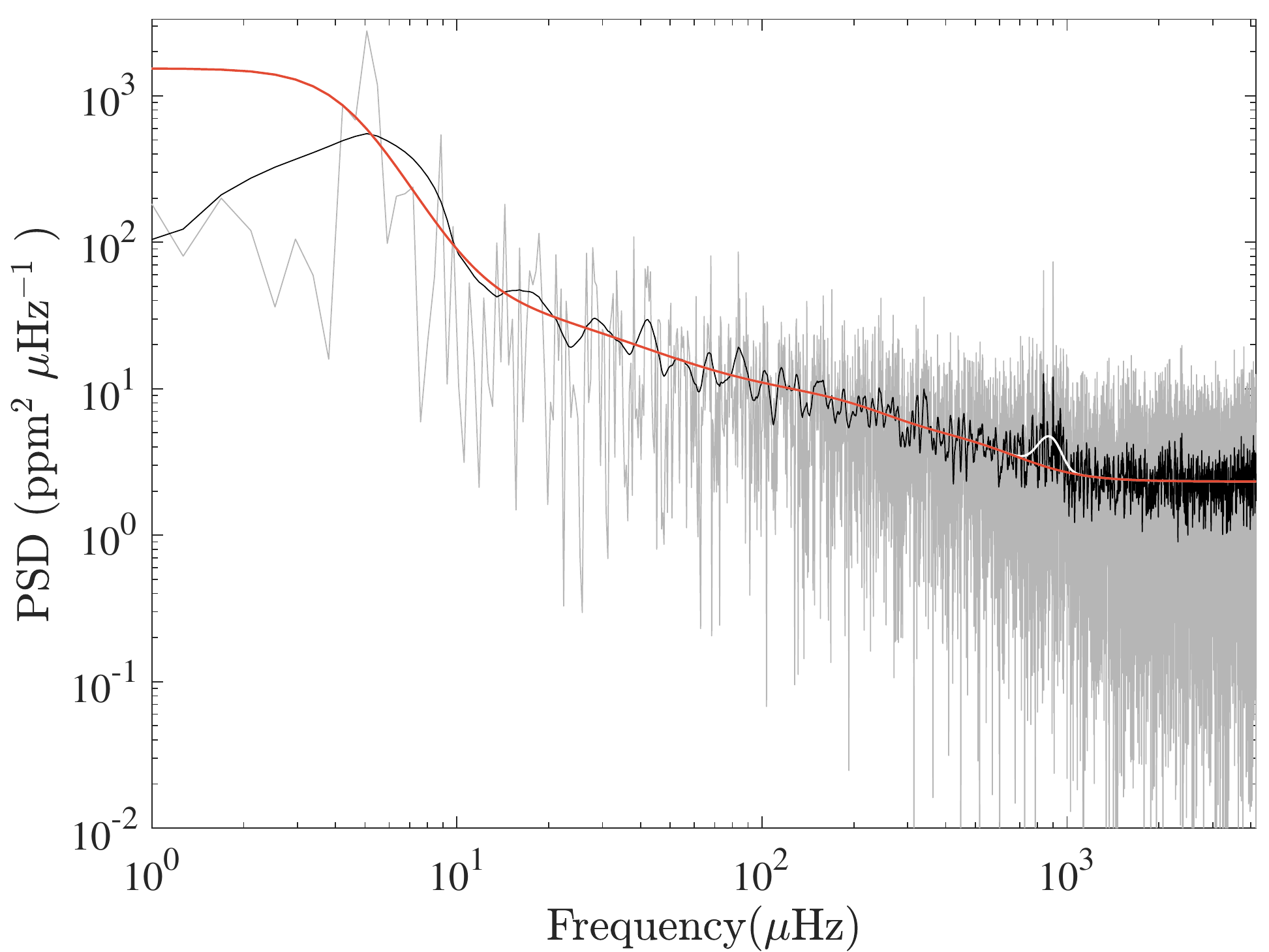} 
 \centering\includegraphics[angle=0,width=\columnwidth,trim=0 0 3 0,clip]{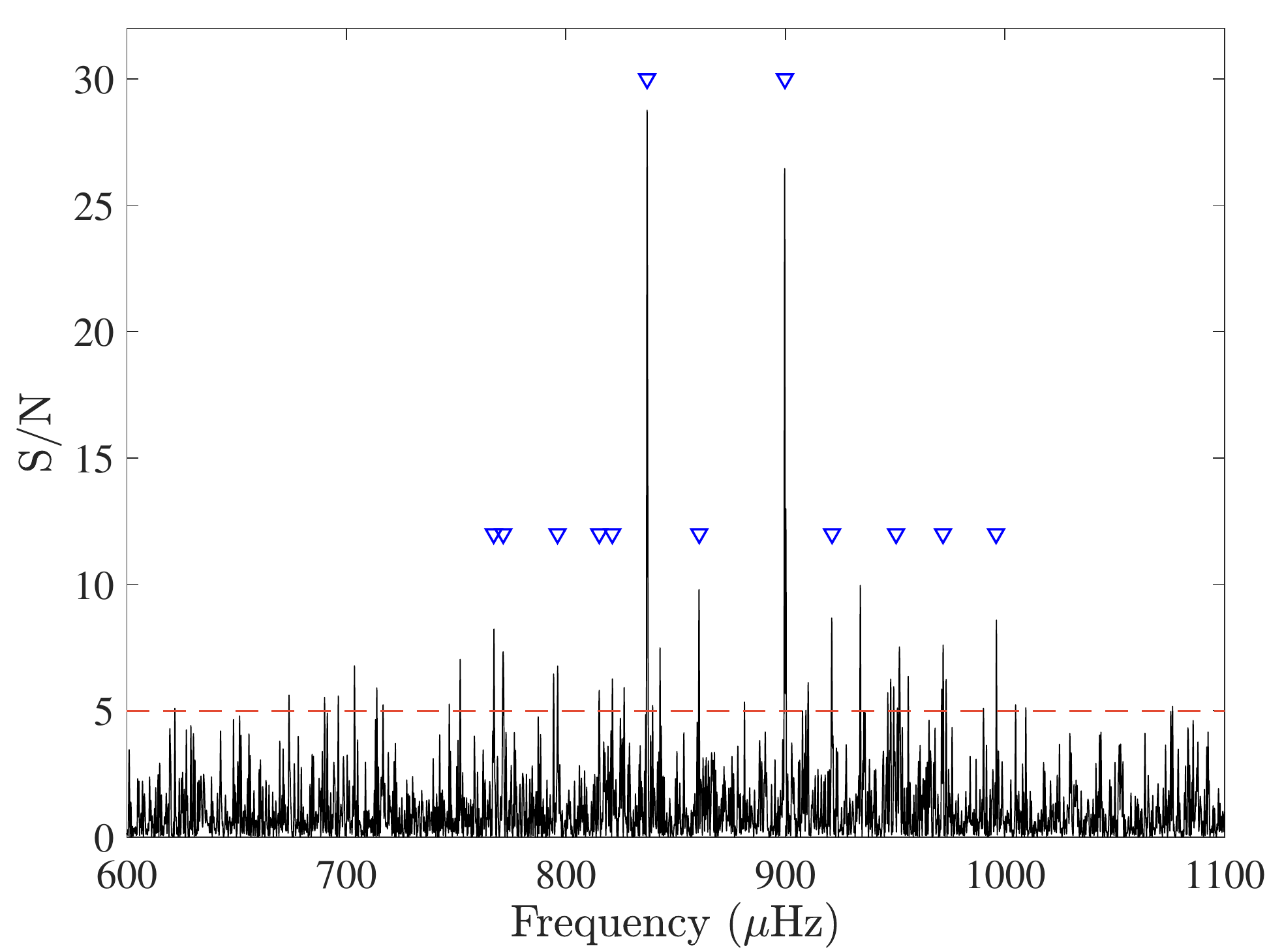} 
 \caption{Top: power spectral density (PSD) of the TESS light curve as a 
function of frequency in $\mu$Hz. The red line represents a fit to the 
background noise, and the white line includes the excess power due to 
oscillations. Bottom: ratio of the power spectrum to the background noise 
as a function of frequency in the range where oscillation modes are 
detected. Within this frequency interval, peaks above the red line have 
$<$\,1\% probability of being due to noise, with the false alarm 
probability decreasing exponentially at higher S/N.\label{fig4}}
 \end{figure}

We compute the power spectral density (PSD) of the TESS light curve (see 
Section~\ref{sec2.2}) to analyze the stellar oscillations. Because the 
signal to noise ratio (S/N) of the oscillations is relatively low, six 
teams analyzed the oscillation spectrum to agree on a frequency set. The 
frequency analysis involves taking into account the background noise 
caused by surface granulation \citep[e.g.,][]{Harvey1988}, then 
extracting the star's eigenmodes (see Figure~\ref{fig4}). Because we are 
dealing with a subgiant star, the latter involves considering mixed 
dipolar ($l=1$) modes, where acoustic waves excited in the convective 
envelope couple with internal gravity waves in the core.

By definition, a mixed mode has a dual character, being both a $p$ mode 
and a $g$ mode. The solution of the continuity equation by 
\cite{Shibahashi1979} led to an implicit expression for the mixed mode 
frequency $\nu_{pg}$ as
\begin{equation}
   \tan\theta_p = q \tan \theta_g,
   \label{shiba}
\end{equation}
where $\theta_p$ and $\theta_g$ are the phase functions of $\nu_{pg}$ 
with respect to the $p$ mode and $g$ mode frequencies, and $q$ is the 
coupling factor. The phase function $\theta_p$ is given by:
\begin{equation}
   \theta_p = \pi \frac{(\nu_{pg} - \nu_{n_{p,1}})}{\Delta\nu(n_p)}
\end{equation}
where 
\begin{equation}
\Delta\nu(n_p) = \left[1 + \alpha(n_p - n_{\rm max})\right] \Delta\nu
\end{equation}
and $\theta_g$ is given by:
\begin{equation}
\theta_g = \pi \frac{1}{P_1} \left(\frac{1}{\nu_{pg}} - \frac{1}{\nu_{n_{g,1}}}\right).
\end{equation}
When there is no coupling, $q = 0$ and there are no mixed modes, so 
$\nu_{pg} = \nu_{n_{p,1}}$. Otherwise, the solutions to Eq.~(\ref{shiba}) 
provide the frequencies of the mixed modes.

Our computation of mixed modes is based upon \cite{Mosser2015}. We first 
assume that the frequencies of radial modes $\nu_{n_p,0}$ are given by:
\begin{equation}
   \nu_{n_p,0} = \left[n_p + \epsilon + \frac{\alpha}{2} (n_p - n_{\rm max})^2\right] \Delta\nu 
   \label{eq_l0_asymp}
\end{equation}
where $n_p$ is the radial order, $\epsilon$ is a phase offset, $\alpha$ 
is the mean curvature of the $l = 0$ pattern as a function of frequency, 
$n_{\rm max}$ is the closest radial order to $\nu_{\rm max}$, and 
$\Delta\nu$ is the mean large frequency separation between consecutive 
radial overtones. The frequencies of the dipolar $p$ modes are given by:
\begin{equation}
   \nu_{n_p,1} = \nu_{n_p,0} + \left(\frac{1}{2} - d_{01}\right) \Delta\nu 
   \label{eq_l1_asymp}
\end{equation}
where $d_{01}$ is the mean separation between $l=0$ and $l=1$ modes of 
a given order $n$, relative to $\Delta\nu$. The periods of dipolar $g$ 
modes $P_{n_g,1}$ are given by:
\begin{equation}
   P_{n_g,1} = (n_g + \alpha_1) P_1
   \label{eq_period_spacing}
\end{equation}
where $n_g$ is the $g$ mode order, $\alpha_1$ is a constant and $P_1$ is 
the period spacing of dipolar modes. The values deduced from 
Eq.~(\ref{shiba}) for 94~Aqr~Aa are $P_1=290.8\pm0.5$~seconds, with 
$q=0.16\pm0.025$. These values are similar to stars with comparable 
$\Delta\nu$ \citep[see][]{Mosser2014}. We used the resulting asymptotic 
$l=1$ mode frequencies as a guide for identification and subsequent 
fitting.

The observed oscillation spectrum is typical of a subgiant star, where 
all $l=1$ modes are of mixed nature, leading to the impression of an 
irregular mode distribution as a function of frequency. Due to the rather 
low S/N, our estimate of the frequency of maximum oscillation power 
suffers from a relatively large uncertainty at $\nu_{\rm max} = 875\pm12\ 
\mu$Hz. Similarly, the determination of $\Delta\nu$ was initially 
ambiguous, with estimates ranging from about 40 to 60 $\mu$Hz. We 
converged to the value $\Delta\nu=50.2\pm0.4\ \mu$Hz, which is 
simultaneously compatible with a fit to both the radial and non-radial 
modes.

From a consensus of the individual teams, we identified four radial 
($l$=0), five dipole ($l$=1), and two quadrupole ($l$=2) modes above our 
significance threshold (S/N\,$\ge5$). The identified frequencies from the 
team that was most representative of the consensus are listed in 
Table~\ref{tab1} and marked with blue triangles in the bottom panel of 
Figure~\ref{fig4}. The two sets of closely-spaced triangles at the lower 
end of the frequency range are pairs of $l$=2 and $l$=0 modes, while the 
two largest peaks are strongly mixed $l$=1 modes. Slightly below the 
frequency range with identified modes, there are a few marginal peaks 
that may be additional $l$=1 and $l$=2 modes for which we did not reach a 
consensus. Finally, there are three peaks above our significance 
threshold that remain unidentified: a relatively strong peak near 
935~$\mu$Hz, a weaker peak adjacent to a mixed mode near 843~$\mu$Hz, and 
a marginal peak close to an $l$=2/$l$=0 pair at 826~$\mu$Hz. Future TESS 
observations in Sector 29 may help to clarify these ambiguities.

 \begin{deluxetable}{ccccc}
 \tablecaption{Identified Oscillation Frequencies.\label{tab1}}
 \tablehead{\colhead{$n$} & \colhead{$l$} & \colhead{$\nu_{nl}^{\rm obs}$ ($\mu$Hz)} & \colhead{$\nu_{nl}^{\rm mod}$ ($\mu$Hz)} & \colhead{$\nu_{nl}^{\rm cor}$ ($\mu$Hz)}}
 \startdata
 15 & 0 & $771.39 \pm 0.19$ & \phantom{1}773.63 & 771.32 \\
 16 & 0 & $821.62 \pm 2.85$ & \phantom{1}826.39 & 821.73 \\
 18 & 0 & $921.29 \pm 0.23$ & \phantom{1}933.78 & 921.93 \\
 19 & 0 & $971.64 \pm 0.29$ & \phantom{1}987.71 & 970.81 \\
 15 & 1 & $794.37 \pm 0.14$ & \phantom{1}801.01 & 797.96 \\
 16 & 1 & $837.02 \pm 0.12$ & \phantom{1}839.83 & 837.48 \\
 16 & 1 & $860.61 \pm 0.14$ & \phantom{1}862.92 & 858.32 \\
 17 & 1 & $899.76 \pm 0.17$ & \phantom{1}909.41 & 900.14 \\
 18 & 1 & $950.49 \pm 1.21$ & \phantom{1}960.71 & 947.10 \\
 19 & 1 & $996.11 \pm 0.15$ &           1010.47 & 994.33 \\
 14 & 2 & $767.21 \pm 0.31$ & \phantom{1}768.71 & 766.60 \\
 15 & 2 & $815.28 \pm 0.15$ & \phantom{1}821.81 & 817.46 
 \enddata
 \end{deluxetable}

\subsection{Modeling the oscillation modes}\label{sec4.2}

To determine the fundamental properties of 94~Aqr~Aa, several teams 
attempted to match the observed oscillation frequencies identified above, 
using stellar evolution models from MESA \citep{Paxton2011}, ASTEC 
\citep{ChristensenDalsgaard2008a}, and the Yale Rotating Evolution Code 
\citep[YREC,][]{Demarque2008} in its non-rotating configuration. We found 
reasonable agreement between the resulting determinations of asteroseismic
radius, mass, and age, with relative dispersions of 2\%, 7\%, and
22\%, respectively. For consistency with our subsequent 
analysis of the angular momentum evolution (see Section~\ref{sec6}), 
below we provide details only for the results obtained with YREC.

We initially constructed a grid of models with masses in the range 
$1.16~M_\odot$ to $1.32~M_\odot$ with a spacing of $0.01~M_\odot$. For 
each mass, models were created with five values of the mixing length 
parameter spanning $\alpha_{\rm MLT}=1.5$ to 2.3, initial helium 
abundances from the primordial helium abundance of 0.248 
\citep{Steigman2010} to 0.30 in steps of 0.01, and initial [Fe/H] in the 
range +0.15 to +0.33 in steps of 0.01. We use the \cite{Grevesse1998} 
solar mixture to convert [Fe/H] to $Z/X$. For each of the parameters, the 
models were evolved from the zero-age main-sequence (ZAMS) to an age of 
11~Gyr. Models were output at intermediate ages.

The models were constructed using OPAL opacities \citep{Iglesias1996} 
supplemented with low temperature opacities from \cite{Ferguson2005}. The 
OPAL equation of state \citep{Rogers2002} was used. All nuclear reaction 
rates are obtained from \cite{Adelberger1998}, except for that of the 
$^{14}N(p,\gamma)^{15}O$ reaction, which we adopt from 
\cite{Formicola2004}. All models included gravitational settling of 
helium and heavy elements using the formulation of \cite{Thoul1994}.  
The oscillation frequencies of the models were calculated with the code 
of \cite{Antia1994}.

The fits to the observations were done in two steps: we first looked for 
models that provided a good match to the frequencies of the $l$=0 and 
$l$=2 modes, in addition to showing consistency with spectroscopic 
constraints \citep[][$T_{\rm eff}=5461\pm40\ {\rm K; [Fe/H]}=+0.23\pm 
0.08$]{Gray2006} and the derived luminosity from Section~\ref{sec2.3} 
($L_{\rm Aa}=3.31^{+0.22}_{-0.07}\ L_\odot$).

The quality of the fit was defined as follows. For each observable, 
$T_{\rm eff}$, [Fe/H] and luminosity $L$, we define a likelihood. For 
instance, the likelihood for effective temperature was defined as 
\begin{equation}
 {\mathcal L}(T_{\rm eff})=D\exp(-\chi^2(T_{\rm eff})/2),
\label{eq:tcal}   
\end{equation}
with
\begin{equation}
\chi^2(T_{\rm eff})=\frac{(T^{\rm obs}_{\rm eff}-T^{\rm mod}_{\rm eff})^2}{\sigma^2_{ T}},
\label{eq:chit}
\end{equation}
where $\sigma_{T}$ is the uncertainty on the effective temperature, and 
$D$ is a normalization constant. We define the likelihoods for [Fe/H] and 
$L$ in a similar manner.

For the frequencies, we first corrected for surface effects using the 
two-term surface correction proposed by \cite{Ball2014}
\begin{eqnarray}
\delta\nu_{nl} & \equiv &\nu_{nl}^{\rm obs}-\nu_{nl}^{\rm mod} \\
 & = &\frac{1}{I_{nl}}\left[ a\left(\frac{\nu_{nl}}{\nu_{\rm ac}}\right)^{-1}
+b\left(\frac{\nu_{nl}}{\nu_{\rm ac}}\right)^3\right],\label{eq:bl}
\end{eqnarray}
where $\delta\nu_{nl}$ is the difference in frequency for a mode of 
degree $l$ and radial order $n$ between the observations and the model, 
$\nu_{nl}$ is the frequency and $I_{nl}$ is the inertia of the mode, and 
$\nu_{\rm ac}$ is the acoustic cut-off frequency, with coefficients $a$ 
and $b$ determined from a generalized least-squares fit to the frequency 
difference of the $l=0$ modes. This allows us to define a likelihood for 
frequencies. We define $\nu^{\rm cor}_{nl}=\nu_{nl}^{\rm mod}-S$, where 
$S$ is defined by the right-hand side of Eq.~(\ref{eq:bl}) but now applied 
to both $l=0$ and $l=2$ modes.
\begin{equation}
\chi^2(\nu)=\frac{(\nu_{nl}^{\rm obs}-\nu_{nl}^{\rm cor})^2}{\sigma^{\rm obs}_{nl}}.
\label{eq:eqchinu}
\end{equation}
Consequently
\begin{equation}
{\mathcal L}(\nu)=C\exp\left(-\frac{\chi^2(\nu)}{2}\right),
\label{eq:nulike}
\end{equation}
where $C$ is a normalization constant.

The total likelihood for each model is then
\begin{equation}
{\mathcal L}_{\rm total}={\mathcal L}(\nu){\mathcal L}(T_{\rm eff}){\mathcal L}([\mathrm{Fe}/\mathrm{H}]){\mathcal L}(L).
\end{equation}
The medians of the marginalized likelihoods of the ensemble of models 
were used to determine the most likely stellar properties, after 
converting them to a probability density by normalizing the likelihood 
by the prior distribution of each parameter.

 \begin{deluxetable}{lccc}
 \tablecaption{Stellar Properties of 94~Aqr~Aa.\label{tab2}}
 \tablehead{\colhead{} & \colhead{Asteroseismic} & \colhead{Other} & \colhead{Source}}
 \startdata
 Radius ($R_\odot$)     & $2.06\pm0.03$  & $2.07\pm0.13$          & (1) \\
 Mass ($M_\odot$)       & $1.22\pm0.03$  & $1.22\pm0.08$          & (2) \\
 Age (Gyr)              & $6.2\pm0.2$    & $6.2^{+0.9}_{-0.7}$    & (3) \\
 $T_{\rm eff}$ (K)      & $5411\pm31$    & $5461\pm40$            & (4) \\
 $[$Fe/H$]$ (dex)       & $+0.15\pm0.05$ & $+0.23\pm0.08$         & (4) \\
 Luminosity ($L_\odot$) & $3.30\pm0.06$  & $3.31^{+0.22}_{-0.07}$ & (5) \\
 $\alpha_{\rm MLT}$     & $1.78\pm0.20$  & $\cdots$               &     \\
 \enddata
 \tablerefs{(1) Section~\ref{sec5.1}; (2) Section~\ref{sec5.2};
 (3) Section~\ref{sec5.3}; (4) \cite{Gray2006}; (5) Section~\ref{sec2.3}}
 \end{deluxetable}

A finer grid in mass and age was created around the most likely values of 
these parameters in order to fit the $l=1$ modes, improving both the
accuracy of the stellar age and the sampling of the posterior distributions.
The likelihoods were 
calculated again, except that now in Eq.~(\ref{eq:eqchinu})\ we also used 
the $l=1$ modes. The optimal asteroseismic properties were derived from 
the new probability density, and they are listed in Table~\ref{tab2} 
along with independent estimates (see Section~\ref{sec5}) and the other 
available constraints. Note that because YREC did not use a global 
optimization technique, it is possible that a better fit to the 
observations exists.\\

\section{Accuracy of the stellar properties}\label{sec5}

 \begin{figure}
 \centering\includegraphics[width=\columnwidth,trim=100 70 80 80,clip]{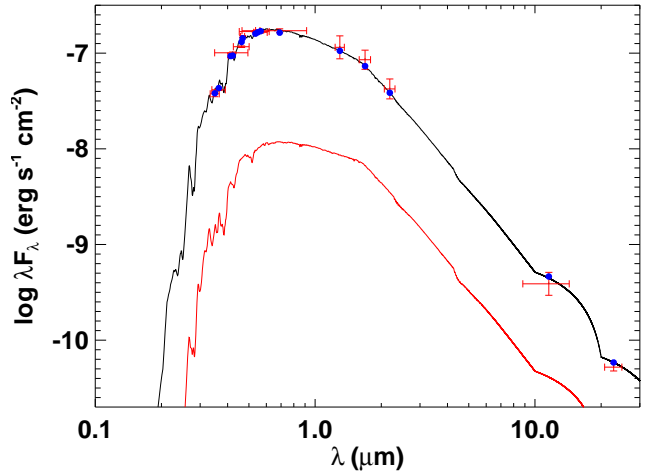}
 \caption{Spectral energy distribution (SED) fit to the broadband 
photometry of 94~Aqr~A. The fit to 94~Aqr~Aa is shown in black, with 
observed fluxes as red symbols and bandpass integrated model fluxes as 
blue symbols. The red curve shows the SED of the companion (Ab), which is 
used to correct for light contamination in the SED of 
94~Aqr~Aa.\label{fig5}}
 \end{figure}

\subsection{Radius from SED}\label{sec5.1}

We performed a fit to the broadband photometry of 94~Aqr~A in order to 
make an independent determination of the stellar radius (see 
Figure~\ref{fig5}). We followed the procedures described in 
\cite{Stassun2016, Stassun2017, Stassun2018}. Briefly, we adopted the 
best available spectroscopic values for $T_{\rm eff}$ and [Fe/H], and 
then fit a standard stellar atmosphere model \citep{Kurucz1992} to the 
broadband spectral energy distribution (SED) in order to determine 
empirically the bolometric flux at Earth ($F_{\rm bol}$). The free 
parameter of the fit was the interstellar extinction. Using the {\it 
Gaia} DR2 parallax, adjusted for the known small systematic offset 
\citep[$-$82~$\mu$as,][]{StassunTorres2018}, the stellar radius was 
then determined via the Stefan-Boltzmann relation.

We adopted the Johnson $UBV$ magnitudes from the \cite{Mermilliod2006} 
homogenized photometric catalog of bright stars, the $B_T V_T$ magnitudes 
from Tycho-2, the Str\"{o}mgren $ubvy$ magnitudes from 
\cite{Paunzen2015}, the $JHK_S$ magnitudes from 2MASS, the W3--W4 
magnitudes from WISE, and the $G$ magnitude from {\it Gaia}. Together, 
the available photometry spans the full stellar SED over the wavelength 
range 0.35--22~$\mu$m.

We adopted the spectroscopic parameters from \cite{Gray2006}, doubling 
the quoted uncertainty on $T_{\rm eff}$ to a more realistic 80~K. We also 
adopted the parameters for the blended Ab component from 
\cite{Docobo2018}, and similarly fit its SED, in order to correct the 
$F_{\rm bol}$ of 94~Aqr~Aa for contamination of light in the broadband 
photometry from the close companion.

The fit has a reduced $\chi^2 = 3.3$ and an extinction of $A_V = 
0.02^{+0.07}_{-0.02}$. The resulting bolometric flux is $F_{\rm bol} = 
(2.18 \pm 0.25) \times 10^{-7}$ erg s$^{-1}$ cm$^{-2}$, which with the 
parallax gives $R_{\rm Aa}=2.07\pm0.13\ R_\odot$, consistent with the 
asteroseismic value ($R = 2.06 \pm 0.03$ $R_\odot$). A similar analysis 
of 94~Aqr~B, adopting the spectroscopic constraints from 
\cite{Fuhrmann2008}, yields $R_{\rm B}=0.85\pm0.03\ R_\odot$.

\subsection{Mass from astrometry and spectroscopy}\label{sec5.2}

\begin{figure}
 \centering\includegraphics[width=\columnwidth,trim=70 70 70 90,clip]{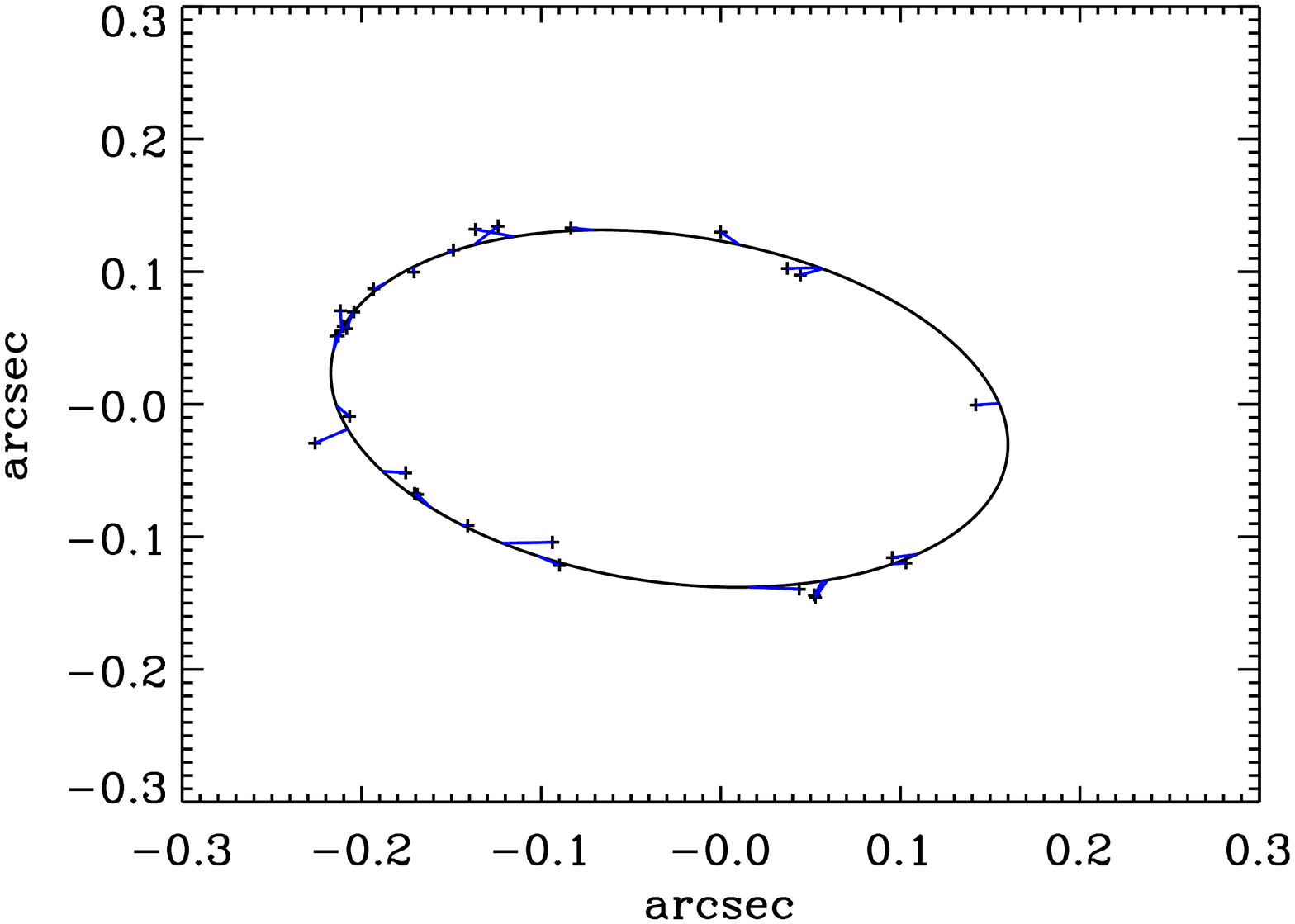} 
 \centering\includegraphics[angle=90,width=\columnwidth,trim=20 16 30 16,clip]{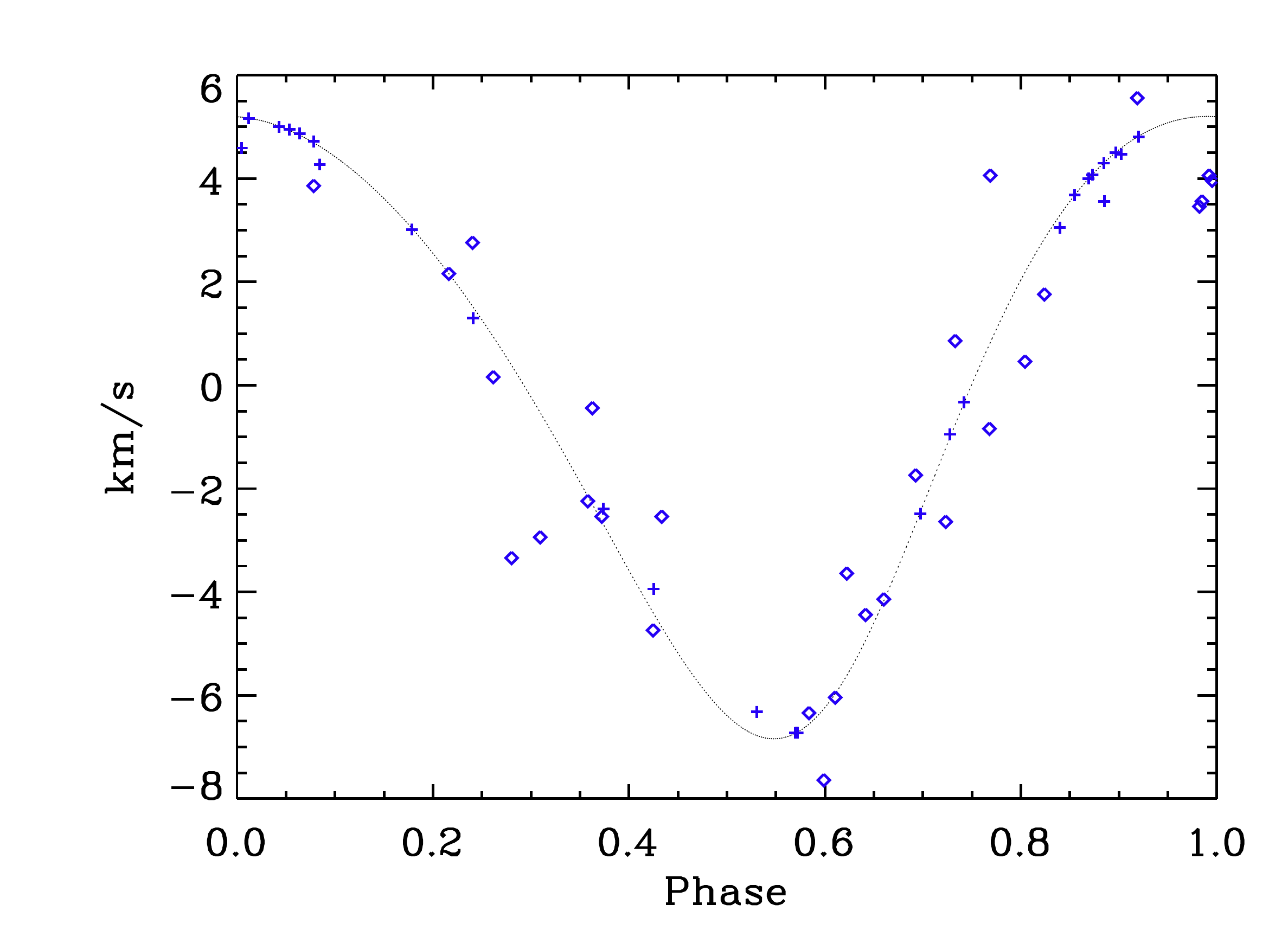}
 \caption{Results of a simultaneous fit to the astrometry (top) and 
spectroscopy (bottom) of 94~Aqr~A. Data are shown with blue points, and 
the fit is shown in black.\label{fig6}}
 \end{figure}

To derive the individual masses of 94~Aqr~Aa and Ab, we used the 
available data from astrometry and spectroscopy. For astrometry, we used 
the same data available to \cite{Docobo2018}. For spectroscopy, we used 
radial velocities measured by \cite{Sarma1962} and \cite{Katoh2013}. We 
then jointly fit the astrometric and spectroscopic data using an MCMC 
approach as described in \cite{Marcadon2018}. The orbital parameters are
given in Table~\ref{tab3}, and the results are illustrated in Figure~\ref{fig6}.

 \begin{deluxetable}{lr}
 \tablecaption{Orbital Parameters of 94~Aqr~A.\label{tab3}}
 \tablehead{\colhead{Parameter~~~~~~~~~~~~~~~~~~~~} & \colhead{Value}}
 \startdata
 Period (years)      & $6.317    \pm 0.005$  \\
 T (year)            & $2012.37  \pm 0.02$   \\
 e                   & $0.1628   \pm 0.0022$ \\
 a (arcsec)          & $0.1925   \pm 0.0032$ \\
 i ($^\circ$)        & $46.72    \pm 1.86$   \\
 $\Omega$ ($^\circ$) & $341.62   \pm 1.11$   \\
 $\omega$ ($^\circ$) & $33.20    \pm 0.88$   \\
 $K$ (km s$^{-1})$   & $6.022    \pm 0.012$  \\
 $\pi$ (mas)         & $44.515   \pm 0.055$  
 \enddata
 \end{deluxetable}

\cite{Docobo2018} used the {\it Gaia} parallax for the Aa/Ab system to 
derive the masses of the two components. Unfortunately, {\it Gaia} DR2 
does not account for binarity. This means the parallax of 94~Aqr~A is not 
useful, because the period of the orbit is about 6.3 years. Instead, we 
used the parallax of 94~Aqr~B \citep[$\pi=44.515\pm0.055$ 
mas;][]{Gaia2018}, which has an orbital period around A that is longer 
than a few hundred years \citep{Mason2001}.

We derived individual masses for Aa and Ab following the same approach as 
\cite{Docobo2018}. The error propagation was done using the MCMC 
solution, thereby taking into account the intrinsic correlations between 
the various orbital parameters. We found $M_{\rm Aa}=1.22 \pm 
0.08\,M_\odot$, in agreement with the asteroseismic value ($M=1.22 \pm 
0.03\,M_\odot$), and $M_{\rm Ab}=0.81 \pm 0.04\,M_\odot$.

\subsection{Age from isochrone fitting}\label{sec5.3}

Because 94~Aqr~Aa is on the subgiant branch and experiencing relatively 
rapid evolution, its age is well constrained from simple isochrone 
fitting. We use the methods described in Section~\ref{sec6} to find a 
best fit stellar model from the observed surface constraints on radius, 
$T_{\rm eff}$, and [Fe/H] in the ``Other'' column of Table~\ref{tab2}. We 
adopt a broad prior on the age of $5.0\pm5.0$~Gyr, while the priors on 
mass, [Fe/H], and mixing length are taken directly from Table~\ref{tab2}. 
We estimate an age of $t_{\rm iso}=6.2^{+0.9}_{-0.7}$~Gyr, which is 
consistent with the asteroseismic value ($t=6.2\pm0.2$~Gyr).

\section{Angular Momentum Evolution}\label{sec6}

By combining the stellar properties from asteroseismology with the 
rotation period determined from the Mount Wilson HK data, we can finally 
model the angular momentum evolution of 94~Aqr~Aa using magnetic braking 
formulations from \citet{vanSaders2013}, \citet{vanSaders2016} and 
\citet{vanSaders2019}. We examined two distinct classes of braking 
models: ``standard" and ``weakened" magnetic braking.

The standard models are described in \citet{vanSaders2013}. We model 
rotational evolution using the \texttt{rotevol} code \citep[as 
in][]{vanSaders2013, vanSaders2016, Somers2016} atop formally 
non-rotating YREC stellar evolution tracks manipulated with the isochrone 
tools in \texttt{kiauhoku} \citep{Claytor2020}. The stellar model grid 
physics are the same as described in \cite{vanSaders2013} with the 
addition of gravitational settling and diffusion, the wider metallicity 
range of \citet{vanSaders2016}, and an Eddington atmosphere. The magnetic 
braking law has four free parameters which are tuned to reproduce the 
observed rotation periods in young open clusters and the Sun: the overall 
normalization of the braking law, $f_k$; the period and duration of the 
disk locking phase $P_{\rm disk}$ and $T_{\rm disk}$, which set the 
initial rotation rate; and the angular rotation velocity at which the 
spin-down transitions from the saturated to unsaturated regime, 
$\omega_{\rm crit}$. For stars near or beyond the end of the 
main-sequence, only $f_k$ is important; the strong dependence of the 
spin-down on rotational velocity means that the late-time evolution is 
insensitive to variations in parameters that affect early-time evolution 
($\omega_{\rm crit}$, $T_{\rm disk}$, $P_{\rm disk}$). In this 
prescription, the spin-down is smooth at late times in much the same 
manner as fully empirical gyrochronology relations 
\citep[e.g.,][]{Barnes2010}.

The weakened braking models are identical to the standard models except 
for one additional free parameter that affects the late-time evolution: a 
critical Rossby number Ro$_{\rm crit}$ beyond which magnetic braking 
ceases and angular momentum is conserved. Due to the weakened braking, 
the modified model generally predicts faster rotation periods than the 
standard model, which allows it to reproduce the rotation periods of old 
asteroseismic calibrator stars \citep{vanSaders2016}. With a fixed $f_k= 
6.6$ (and $\omega_{\rm crit} = 3.4 \times 10^{-5} $ s$^{-1}$, $P_{\rm 
disk} = 8.1$, $T_{\rm disk} = 0.28$), \cite{vanSaders2016} found Ro$_{\rm 
crit} = 2.16$ when calibrating against 21 {\it Kepler} asteroseismic 
targets.

\subsection{Rotational modeling of 94~Aqr~Aa}\label{sec6.1}

For 94~Aqr~Aa, we used the asteroseismically constrained properties to 
predict the observed rotation period under the two different braking 
prescriptions. In contrast to the main-sequence, envelope expansion on 
the subgiant branch (SGB) increasingly dominates over magnetic braking in 
the rotational evolution as the star approaches the base of the giant 
branch. This means that the predicted rotation is tied strongly to age, 
but also to both HR diagram position and stellar mass.

We used a Monte Carlo approach \citep[{\tt emcee},][]{ForemanMackey2013} 
to search our model grids in mass, age, bulk composition (Z/X), and 
mixing length ($\alpha_{\rm MLT}$) to match the best fit values of 
$T_{\rm eff}$, [Fe/H], and radius from the ``Asteroseismic'' column in 
Table \ref{tab2}. We adopted Gaussian priors on the mass, age, [Fe/H], 
and $\alpha_{\rm MLT}$ centered on $1.22\,M_{\odot}$, 6.2~Gyr, +0.15~dex, 
and 1.78 with 1$\sigma$ widths of $0.03\,M_{\odot}$, 0.2~Gyr, 0.05~dex, 
and 0.2, respectively. A total of 8 chains were run for 100000 steps 
each, with the first 5000 steps discarded as burn-in. Such a run 
corresponds to $>$\,1000 autocorrelation times in all variables of 
interest.

\begin{figure}[t]
 \centering\includegraphics[width=\columnwidth,trim=7 7 0 0,clip]{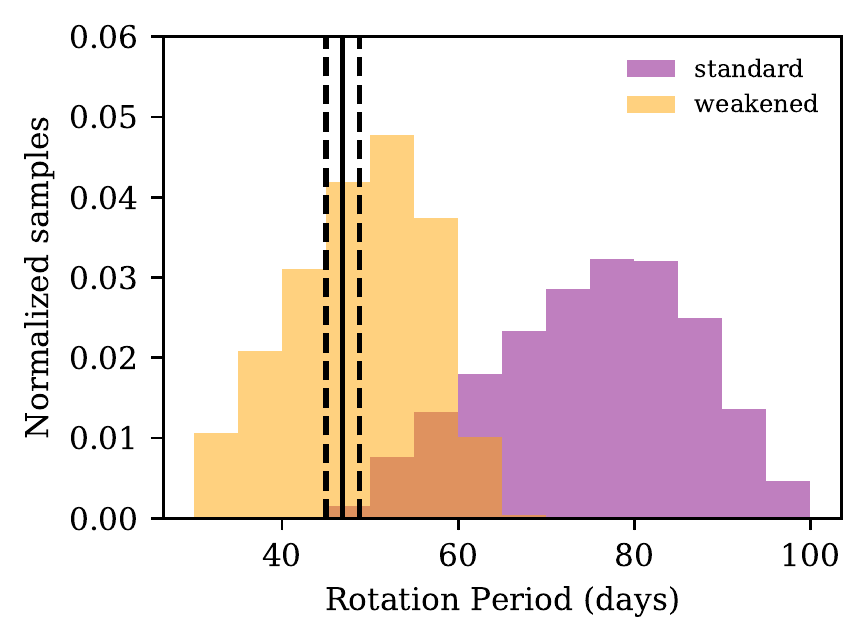}
 \caption{Predictions from a standard spin-down model (purple) and 
weakened braking model (orange) for the rotation period of 94~Aqr~Aa. The 
observed rotation period from Mount Wilson data is shown with black 
vertical lines.\label{fig7}}
 \end{figure}

We predict a rotation period for the standard braking model of $P = 
76^{+11}_{-13}$~days with the asteroseismic surface constraints. For a 
weakened braking model with Ro$_{\rm crit}$, we predict $P = 
49^{+7}_{-9}$~days, in good agreement with the observed period of 
$46.9\pm1.9$~days (see Figure \ref{fig7}). If instead we adopt the 
spectroscopic values for $T_{\rm eff}$ and [Fe/H], we would predict $P = 
63^{+12}_{-11}$~days for the standard model, and $P = 41\pm8$~days for 
weakened braking. In both cases the weakened braking model provides 
better agreement with the observed rotation.

\subsection{Robustness to assumptions}\label{sec6.2}

There are two types of physical assumptions that can affect our 
conclusions: assumptions about the magnetic braking and angular momentum 
transport itself, and physical uncertainties in the underlying stellar 
models. We argue that the latter are of most concern for the weakened 
braking interpretation.

We have made a host of assumptions about the braking, most of which we 
expect cannot shift the standard model prediction toward the shorter 
observed rotation period. We've assumed a single set of initial 
conditions despite the fact that stars in nature display a range of birth 
rotation rates, but even if we had launched our best fit models rotating 
at breakup velocity on the ZAMS (3 times faster than assumed in our 
models, which have a ZAMS period of less than 1 day) 
the predicted period for the standard model would differ by only 
1 day. We haven't allowed for differential rotation, but for cases in 
which the core rotationally decouples from the envelope 
\citep[e.g.][]{Denissenkov2010, MacGregor1991} we expect slower observed 
envelope rotation rates, not faster \citep{Deheuvels2014}. If the star 
had a strong radial differential rotation on the main-sequence it could 
result in faster-than-expected rotation on the SGB, but indications are 
that main-sequence and early SGB stars have little radial differential 
rotation \citep[e.g.][]{Thompson1996, Saio2015, Benomar2015, 
Deheuvels2020}. 94~Aqr~Aa is not a single star, but its 6.3 year orbit is 
sufficiently wide that we expect it to behave as a single star in terms 
of rotational evolution.

By contrast, reasonable changes to the underlying stellar physics have 
the ability to shift the predicted rotation period in the standard model 
by more than 10 days. The asteroseismic fits in Section~\ref{sec4.2} 
allowed the mixing length to be a free parameter, rather than fixing it 
to the solar value as is the norm in most non-seismic analyses. We have 
found that our results are sensitive to the choice of mixing length.

Smaller values of $\alpha_{\rm MLT}$ yield models with shorter periods. 
Changing the mixing length tends to shift subgiant tracks along the SGB, 
meaning that constraints in temperature, radius, and luminosity can be 
matched with a model of essentially the same stellar mass. However, at a 
fixed location in the HR diagram, models of the same mass but different 
mixing lengths will have different convective overturn timescales, 
affecting the predicted rotation period. A $1.2\,M_{\odot}$, 0.2~dex 
model at $2.0\,R_{\odot}$ with a mixing length of 1.8 has a period of 60 
days, but the same model with a mixing length of 1.6 has a period of 46 
days at $2.0\,R_{\odot}$--- a difference comparable to that between 
period predictions for a standard and weakened model of magnetic braking. 
While we have allowed the mixing length to vary in our fit and 
incorporated the (fairly broad) asteroseismic prior on $\alpha_{\rm 
MLT}$, the sensitivity of the period to the choice of mixing length means 
that we cannot rule out the possibility that the tension between the 
observed and predicted period for standard braking arises from an 
inappropriate choice of the mixing length.

Efforts to quantify how the mixing length should vary as a function of 
stellar properties have yielded conflicting results. Observational 
estimates from asteroseismology \citep{bonaca2012, Creevey2017, 
tayar2017, viani2018} find that the mixing length should increase as the 
metallicity increases, while simulations of convection generally arrive 
at the opposite conclusion \citep{magic2015}. \citet{viani2018} predicts 
that a star with the surface gravity, temperature and metallicity of 
94~Aqr~Aa should have a super-solar mixing length for any of the 
relations they provide. Larger mixing lengths result in longer rotation 
periods, and would reinforce the tension between the standard model and 
the observed rotation. We note that, in contrast to these predictions, 
the asteroseismic mixing length for 94~Aqr~Aa in Table~\ref{tab2} is 
sub-solar ($\alpha_{\rm MLT,\odot}=1.98$).

The apparent need for weakened braking could also be spurious if our 
stellar mass is underestimated. Rotation rate is a very strong function 
of mass, particularly near the \cite{Kraft1967} break, where stars above 
the break in mass rotate rapidly, while less massive stars rotate more 
slowly. The dichotomy is a result of the diminishing convective envelopes 
in more massive stars, and the consequently weak large-scale fields and 
magnetic braking. If we fix the mixing length to $\alpha_{\rm MLT} = 
1.78$ and broaden the priors on mass, age, and metallicity to $1\sigma$ 
Gaussian widths of $0.5\,M_{\odot}$, 2.0~Gyr, and 0.2~dex, the resulting 
increase of $0.05\,M_{\odot}$ in mass is sufficient to shift the 
predicted rotation period from $\sim$71 days to $55$ days. However, we 
note that changing the mixing length to $\sim$1.5 can match the observed 
rotation period with no upward adjustment to the mass, so we consider the 
poorly constrained mixing length to be the more intractable source of 
uncertainty.

These issues are connected to challenges inherent in stellar modeling 
that have stood for decades, and the solution is unlikely to emerge from 
this article. In the following section, we explore the possibility that 
94~Aqr~Aa has experienced weakened magnetic braking, but we caution the 
reader about the sizable caveats to this interpretation that we have 
outlined above.

\newpage
\section{Summary and Discussion}\label{sec7}

By combining Mount Wilson observations of magnetic variability with TESS 
asteroseismic measurements of the G8 subgiant 94~Aqr~Aa, we have 
discovered new evidence for weakened magnetic braking 
\citep{vanSaders2016} and the possibility of a ``born-again'' dynamo in 
evolved stars, as we discuss below.

A reanalysis of 35 years of HK observations (Section~\ref{sec3}) yielded 
rotation and magnetic activity cycle periods for 94~Aqr~Aa (P$^{\rm 
Aa}_{\rm rot} = 46.9 \pm 1.9$~days, P$^{\rm Aa}_{\rm cyc} = 19.35 \pm 
0.18$~yr) and a cycle period for 94~Aqr~B (P$^{\rm B}_{\rm cyc} = 9.13 
\pm 0.03$~yr). The amplitude of the observed variability in the blended A 
component allowed us to attribute these properties to the subgiant (Aa) 
because it contributes 97\% of the light in the relevant bandpasses.

Asteroseismology of 94~Aqr~Aa from TESS observations (Section~\ref{sec4}) 
yielded precise determinations of the stellar radius ($R = 2.06 \pm 0.03\ 
R_\odot$), mass ($M = 1.22 \pm 0.03\ M_\odot$) and age ($t = 6.2 \pm 
0.2$~Gyr). We established the absolute accuracy of these properties 
(Section~\ref{sec5}) with independent estimates of the stellar radius 
from SED fitting ($R_{\rm Aa} = 2.07 \pm 0.13\ R_\odot$), the stellar 
mass from a close binary orbit ($M_{\rm Aa} = 1.22 \pm 0.08\ M_\odot$), 
and the age from isochrone fitting ($t_{\rm iso} = 
6.2^{+0.9}_{-0.7}$~Gyr).

Using the asteroseismic properties from Section~\ref{sec4}, we attempted 
to reproduce the observed rotation period from Section~\ref{sec3} with 
angular momentum evolution models (Section~\ref{sec6}) that adopted 
either standard spin-down or the weakened magnetic braking proposed by 
\cite{vanSaders2016}. The standard model predicts a rotation period ($P = 
76^{+11}_{-13}$~days) that is substantially longer than suggested by the 
observations, while the model with weakened magnetic braking ($P = 
49^{+7}_{-9}$~days) more closely reproduces the observed rotation period 
with stalled spin-down at a critical Rossby number Ro$_{\rm crit}=2.16$ 
\citep{vanSaders2016}. Note that with these models, the Rossby
number of the Sun is Ro$_\odot\sim2.2$, comparable to the critical value.

The fact that the G8 subgiant shows a magnetic activity cycle provides an 
interesting constraint on stellar dynamo models. According to the 
scenario proposed by \cite{Metcalfe2017}, activity cycles should 
gradually grow longer with the rotation period along the two sequences 
identified by \cite{BohmVitense2007}. When a star reaches the critical 
Rossby number suggested by \cite{vanSaders2016}, the rotation period 
remains relatively constant while the activity cycle appears to grow 
longer and weaker before disappearing entirely. The resulting ``flat 
activity'' star still shows magnetic activity on small scales, allowing 
rotation periods to be measured, but the mean activity level is 
approximately constant on longer timescales. Such stars have previously 
been interpreted as Maunder minimum candidates \citep{Judge2004}, but at 
least some of them may be the end-states of large-scale stellar dynamos.

 \begin{figure}[!t]
 \centering\includegraphics[width=\columnwidth,trim=7 7 0 0,clip]{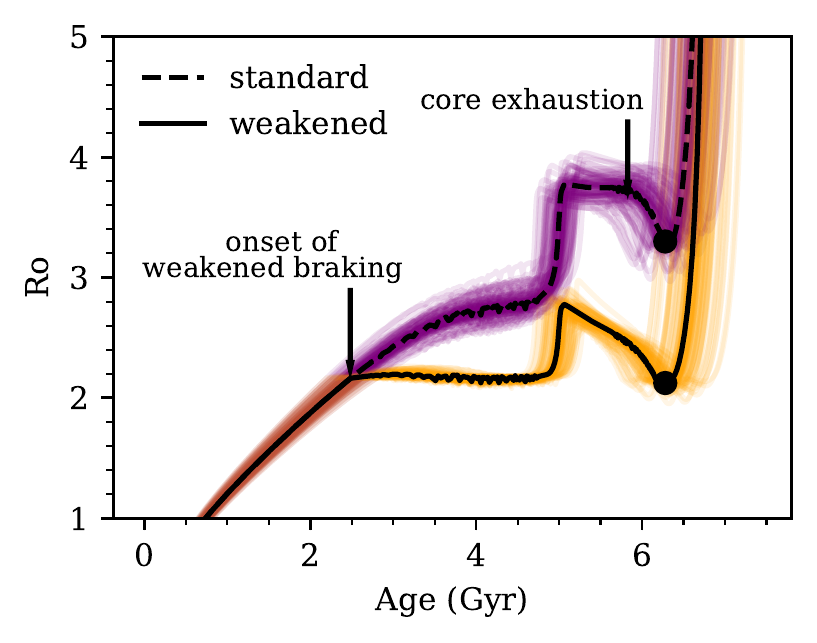}
 \caption{Predictions from a standard spin-down model (purple/dashed) and 
weakened braking model (orange/solid) for the evolution of the Rossby 
number in 94~Aqr~Aa. For each model, the best-fit combination of mass and 
composition is shown as a dark line and the best fit age is marked with a 
solid point. In addition, 250 randomly drawn posterior samples of mass 
and composition are shown, truncated at the end of the subgiant 
branch.\label{fig8}}
 \end{figure} 

If the critical Rossby number represents a threshold above which 
large-scale dynamos no longer operate \citep{Tripathi2018}, then models 
with weakened magnetic braking may help to explain the existence of a 
cycle in 94~Aqr~Aa. The mass of this G8 subgiant suggests that it evolved 
from an F-type star on the main-sequence. As such, it may have had a 
relatively short activity cycle until it reached the critical Rossby 
number after about 2.5~Gyr (see Figure~\ref{fig8}) at a rotation period 
near 15~days \citep[e.g., see][their Fig.~1]{Metcalfe2019b}. The cycle 
would have then grown longer and weaker for about 2~Gyr at nearly 
constant Rossby number. When hydrogen core-burning ceased, the core would 
have contracted and the star would become hotter with a thinner 
convection zone, pushing it above Ro$_{\rm crit}$ and making it a ``flat 
activity'' star. However, the star would subsequently expand and cool 
when hydrogen shell-burning began, slowing its rotation through 
conservation of angular momentum and deepening the outer convection zone. 
For a small range of masses above the solar value, these evolutionary 
effects (solid line in Figure~\ref{fig8}) can push the Rossby number back 
below Ro$_{\rm crit}$ so the star can reinvigorate large-scale dynamo 
action and briefly sustain an activity cycle before ascending the red 
giant branch. This scenario for a ``born-again'' dynamo is simply not 
possible with standard models (dashed line in Figure~\ref{fig8}).

A similar mechanism may help explain the existence of magnetic cycles in 
subgiants that evolved from more massive F-type stars, which had never 
previously sustained a large-scale dynamo. For example, 
\cite{Egeland2018} used Mount Wilson and Lowell observations to identify 
an activity cycle in the subgiant component of the HD\,81809 system. The 
mass of this star ($1.58\pm0.26\ M_\odot$) would place it above the Kraft 
break on the main-sequence, without a substantial outer convection zone 
to help build a large-scale magnetic field. Consequently, it would not 
have experienced significant magnetic braking during its main-sequence 
lifetime, and it would only have slowed to its current rotation period 
($40.2\pm2.3$~days) through expansion on the subgiant branch. The deeper 
convection zone during this evolutionary phase could finally support 
large-scale dynamo action for the first time in its life, explaining the 
observed activity cycle.

The results presented above demonstrate the power of combining magnetic 
variability data from Mount Wilson and other programs with new 
asteroseismic observations from TESS. With accurate determinations of the 
basic stellar properties such as radius, mass, and age, we can finally 
reveal the evolutionary threads that connect stars with known rotation 
rates and magnetic activity cycles. Over the coming years, this approach 
promises to yield additional insights about magnetic stellar evolution, 
particularly beyond the middle of main-sequence lifetimes.

\section*{Acknowledgments}

T.S.M.~acknowledges support from NSF grant AST-1812634, NASA grants NNX16AB97G and 80NSSC20K0458, and a Visiting Fellowship at the Max Planck Institute for Solar System Research. Computational time at the Texas Advanced Computing Center was provided through XSEDE allocation TG-AST090107. 
J.v.S. acknowledges support from NASA through the TESS Guest Investigator Program (80NSSC18K1584). 
S.B.~acknowledges NASA grant 80NSSC19K0374. 
D.B.~acknowledges support from NASA through the Living With A Star Program (NNX16AB76G) and from the TESS GI Program under awards 80NSSC18K1585 and 80NSSC19K0385. 
W.J.C., W.H.B.~and M.B.N.~acknowledge support from the UK Space Agency. 
Funding for the Stellar Astrophysics Centre is provided by The Danish National Research Foundation (Grant agreement no.: DNRF106). 
R.E.~was supported by the NCAR Advanced Study Program Postdoctoral Fellowship. 
The National Center for Atmospheric Research is sponsored by the National Science Foundation. 
R.A.G.~and B.M.~acknowledge support from the CNES PLATO grant. 
P.G.~acknowledges funding from the German Aerospace Center (Deutsches Zentrum f\"ur Luft- und Raumfahrt) under PLATO Data Center grant 50OO1501. 
D.H.~acknowledges support from NASA through the TESS Guest Investigator Program (80NSSC18K1585, 80NSSC19K0379). 
T.R.~acknowledges support from the European Research Council (ERC) under the European Union's Horizon 2020 research and innovation programme (grant agreement No. 715947). 
T.A.~acknowledges support from the Programme de Physique Stellaire et Plan\'etaire. 
T.R.B.~acknowledges support from the Australian Research Council. 
L.G.C.~acknowledges support from grant FPI-SO from the Spanish Ministry of Economy and Competitiveness (MINECO) (research project SEV-2015-0548-17-2 and predoctoral contract BES-2017-082610). 
A.J.~Acknowledges support from the State Research Agency (AEI) of the Spanish Ministry of Science, Innovation and Universities (MCIU). 
H.K.~acknowledges support from the European Social Fund via the Lithuanian Science Council (LMTLT) grant No.~09.3.3-LMT-K-712-01-0103.
M.N.L.~acknowledges support from the ESA PRODEX programme. 
S.M.~acknowledges support from the Spanish Ministry with the Ramon y Cajal fellowship number RYC-2015-17697. 
Z.\c{C}.O., S.\"O.~and M.Y.~acknowledge support from the Scientific and Technological Research Council of Turkey (T\"UB\.ITAK:118F352). 
A.R.G.S.~acknowledges support from NASA grant NNX17AF27G.
This work benefited from discussions within the international team ``The Solar and Stellar Wind Connection: Heating processes and angular momentum loss'' at the International Space Science Institute (ISSI). 



\begin{thebibliography}{}
\expandafter\ifx\csname natexlab\endcsname\relax\def\natexlab#1{#1}\fi
\providecommand{\url}[1]{\href{#1}{#1}}
\providecommand{\dodoi}[1]{doi:~\href{http://doi.org/#1}{\nolinkurl{#1}}}
\providecommand{\doeprint}[1]{\href{http://ascl.net/#1}{\nolinkurl{http://ascl.net/#1}}}
\providecommand{\doarXiv}[1]{\href{https://arxiv.org/abs/#1}{\nolinkurl{https://arxiv.org/abs/#1}}}

\bibitem[{{Adelberger} {et~al.}(1998){Adelberger}, {Austin}, {Bahcall},
  {Balantekin}, {Bogaert}, {Brown}, {Buchmann}, {Cecil}, {Champagne}, {de
  Braeckeleer}, {Duba}, {Elliott}, {Freedman}, {Gai}, {Goldring}, {Gould},
  {Gruzinov}, {Haxton}, {Heeger}, {Henley}, {Johnson}, {Kamionkowski},
  {Kavanagh}, {Koonin}, {Kubodera}, {Langanke}, {Motobayashi}, {Pandharipande},
  {Parker}, {Robertson}, {Rolfs}, {Sawyer}, {Shaviv}, {Shoppa}, {Snover},
  {Swanson}, {Tribble}, {Turck-Chi{\`e}ze}, \& {Wilkerson}}]{Adelberger1998}
{Adelberger}, E.~G., {Austin}, S.~M., {Bahcall}, J.~N., {et~al.} 1998, Reviews
  of Modern Physics, 70, 1265, \dodoi{10.1103/RevModPhys.70.1265}

\bibitem[{{Antia} \& {Basu}(1994)}]{Antia1994}
{Antia}, H.~M., \& {Basu}, S. 1994, \aaps, 107, 421

\bibitem[{{Baliunas} {et~al.}(1996){Baliunas}, {Sokoloff}, \&
  {Soon}}]{Baliunas1996}
{Baliunas}, S., {Sokoloff}, D., \& {Soon}, W. 1996, \apjl, 457, L99,
  \dodoi{10.1086/309891}

\bibitem[{{Baliunas} {et~al.}(1983){Baliunas}, {Hartmann}, {Noyes}, {Vaughan},
  {Preston}, {Frazer}, {Lanning}, {Middelkoop}, \& {Mihalas}}]{Baliunas1983}
{Baliunas}, S.~L., {Hartmann}, L., {Noyes}, R.~W., {et~al.} 1983, \apj, 275,
  752, \dodoi{10.1086/161572}

\bibitem[{{Baliunas} {et~al.}(1995){Baliunas}, {Donahue}, {Soon}, {Horne},
  {Frazer}, {Woodard-Eklund}, {Bradford}, {Rao}, {Wilson}, {Zhang}, {Bennett},
  {Briggs}, {Carroll}, {Duncan}, {Figueroa}, {Lanning}, {Misch}, {Mueller},
  {Noyes}, {Poppe}, {Porter}, {Robinson}, {Russell}, {Shelton}, {Soyumer},
  {Vaughan}, \& {Whitney}}]{Baliunas1995}
{Baliunas}, S.~L., {Donahue}, R.~A., {Soon}, W.~H., {et~al.} 1995, \apj, 438,
  269, \dodoi{10.1086/175072}

\bibitem[{{Ball} \& {Gizon}(2014)}]{Ball2014}
{Ball}, W.~H., \& {Gizon}, L. 2014, \aap, 568, A123

\bibitem[{{Barnes}(2010)}]{Barnes2010}
{Barnes}, S.~A. 2010, \apj, 722, 222, \dodoi{10.1088/0004-637X/722/1/222}

\bibitem[{{Benomar} {et~al.}(2015){Benomar}, {Takata}, {Shibahashi},
  {Ceillier}, \& {Garc{\'\i}a}}]{Benomar2015}
{Benomar}, O., {Takata}, M., {Shibahashi}, H., {Ceillier}, T., \&
  {Garc{\'\i}a}, R.~A. 2015, \mnras, 452, 2654, \dodoi{10.1093/mnras/stv1493}

\bibitem[{{B{\"o}hm-Vitense}(2007)}]{BohmVitense2007}
{B{\"o}hm-Vitense}, E. 2007, \apj, 657, 486, \dodoi{10.1086/510482}

\bibitem[{{Bonaca} {et~al.}(2012){Bonaca}, {Tanner}, {Basu}, {Chaplin},
  {Metcalfe}, {Monteiro}, {Ballot}, {Bedding}, {Bonanno}, {Broomhall},
  {Bruntt}, {Campante}, {Christensen-Dalsgaard}, {Corsaro}, {Elsworth},
  {Garc{\'\i}a}, {Hekker}, {Karoff}, {Kjeldsen}, {Mathur}, {R{\'e}gulo},
  {Roxburgh}, {Stello}, {Trampedach}, {Barclay}, {Burke}, \&
  {Caldwell}}]{bonaca2012}
{Bonaca}, A., {Tanner}, J.~D., {Basu}, S., {et~al.} 2012, \apjl, 755, L12,
  \dodoi{10.1088/2041-8205/755/1/L12}

\bibitem[{{Brandenburg} {et~al.}(2017){Brandenburg}, {Mathur}, \&
  {Metcalfe}}]{Brandenburg2017}
{Brandenburg}, A., {Mathur}, S., \& {Metcalfe}, T.~S. 2017, \apj, 845, 79,
  \dodoi{10.3847/1538-4357/aa7cfa}

\bibitem[{{Buzasi} {et~al.}(2016){Buzasi}, {Lezcano}, \&
  {Preston}}]{Buzasi2016}
{Buzasi}, D., {Lezcano}, A., \& {Preston}, H.~L. 2016, Journal of Space Weather
  and Space Climate, 6, A38, \dodoi{10.1051/swsc/2016033}

\bibitem[{{Buzasi} {et~al.}(2015){Buzasi}, {Carboneau}, {Hessler}, {Lezcano},
  \& {Preston}}]{Buzasi2015}
{Buzasi}, D.~L., {Carboneau}, L., {Hessler}, C., {Lezcano}, A., \& {Preston},
  H. 2015, in IAU General Assembly, Vol.~29, 2256843

\bibitem[{{Casagrande} {et~al.}(2014){Casagrande}, {Portinari}, {Glass},
  {Laney}, {Silva Aguirre}, {Datson}, {Andersen}, {Nordstr{\"o}m}, {Holmberg},
  {Flynn}, \& {Asplund}}]{Casagrande2014}
{Casagrande}, L., {Portinari}, L., {Glass}, I.~S., {et~al.} 2014, \mnras, 439,
  2060, \dodoi{10.1093/mnras/stu089}

\bibitem[{{Christensen-Dalsgaard}(2008)}]{ChristensenDalsgaard2008a}
{Christensen-Dalsgaard}, J. 2008, \apss, 316, 13,
  \dodoi{10.1007/s10509-007-9675-5}

\bibitem[{{Claytor} {et~al.}(2020){Claytor}, {van Saders}, {Santos},
  {Garc{\'\i}a}, {Mathur}, {Tayar}, {Pinsonneault}, \&
  {Shetrone}}]{Claytor2020}
{Claytor}, Z.~R., {van Saders}, J.~L., {Santos}, {\^A}. R.~G., {et~al.} 2020,
  \apj, 888, 43, \dodoi{10.3847/1538-4357/ab5c24}

\bibitem[{{Creevey} {et~al.}(2017){Creevey}, {Metcalfe}, {Schultheis},
  {Salabert}, {Bazot}, {Th{\'e}venin}, {Mathur}, {Xu}, \&
  {Garc{\'\i}a}}]{Creevey2017}
{Creevey}, O.~L., {Metcalfe}, T.~S., {Schultheis}, M., {et~al.} 2017, \aap,
  601, A67, \dodoi{10.1051/0004-6361/201629496}

\bibitem[{{Deheuvels} {et~al.}(2020){Deheuvels}, {Ballot}, {Eggenberger},
  {Spada}, {Noll}, \& {den Hartogh}}]{Deheuvels2020}
{Deheuvels}, S., {Ballot}, J., {Eggenberger}, P., {et~al.} 2020, arXiv
  e-prints, arXiv:2007.02585.
\newblock \doarXiv{2007.02585}

\bibitem[{{Deheuvels} \& {Michel}(2011)}]{Deheuvels2011}
{Deheuvels}, S., \& {Michel}, E. 2011, \aap, 535, A91,
  \dodoi{10.1051/0004-6361/201117232}

\bibitem[{{Deheuvels} {et~al.}(2014){Deheuvels}, {Do{\u{g}}an}, {Goupil},
  {Appourchaux}, {Benomar}, {Bruntt}, {Campante}, {Casagrande}, {Ceillier},
  {Davies}, {De Cat}, {Fu}, {Garc{\'\i}a}, {Lobel}, {Mosser}, {Reese},
  {Regulo}, {Schou}, {Stahn}, {Thygesen}, {Yang}, {Chaplin},
  {Christensen-Dalsgaard}, {Eggenberger}, {Gizon}, {Mathis},
  {Molenda-{\.Z}akowicz}, \& {Pinsonneault}}]{Deheuvels2014}
{Deheuvels}, S., {Do{\u{g}}an}, G., {Goupil}, M.~J., {et~al.} 2014, \aap, 564,
  A27, \dodoi{10.1051/0004-6361/201322779}

\bibitem[{{Demarque} {et~al.}(2008){Demarque}, {Guenther}, {Li}, {Mazumdar}, \&
  {Straka}}]{Demarque2008}
{Demarque}, P., {Guenther}, D.~B., {Li}, L.~H., {Mazumdar}, A., \& {Straka},
  C.~W. 2008, \apss, 316, 31, \dodoi{10.1007/s10509-007-9698-y}

\bibitem[{{Denissenkov} {et~al.}(2010){Denissenkov}, {Pinsonneault},
  {Terndrup}, \& {Newsham}}]{Denissenkov2010}
{Denissenkov}, P.~A., {Pinsonneault}, M., {Terndrup}, D.~M., \& {Newsham}, G.
  2010, \apj, 716, 1269, \dodoi{10.1088/0004-637X/716/2/1269}

\bibitem[{{Docobo} {et~al.}(2018){Docobo}, {Tamazian}, {Campo}, \&
  {Piccotti}}]{Docobo2018}
{Docobo}, J.~A., {Tamazian}, V.~S., {Campo}, P.~P., \& {Piccotti}, L. 2018,
  \aj, 156, 85, \dodoi{10.3847/1538-3881/aad179}

\bibitem[{{Donahue} {et~al.}(1996){Donahue}, {Saar}, \&
  {Baliunas}}]{Donahue1996}
{Donahue}, R.~A., {Saar}, S.~H., \& {Baliunas}, S.~L. 1996, \apj, 466, 384,
  \dodoi{10.1086/177517}

\bibitem[{{Egeland}(2017)}]{Egeland2017}
{Egeland}, R. 2017, PhD thesis, Montana State University, Bozeman, Montana, USA

\bibitem[{{Egeland}(2018)}]{Egeland2018}
---. 2018, \apj, 866, 80, \dodoi{10.3847/1538-4357/aadf86}

\bibitem[{{Egeland} {et~al.}(2017){Egeland}, {Soon}, {Baliunas}, {Hall},
  {Pevtsov}, \& {Bertello}}]{Egeland2017b}
{Egeland}, R., {Soon}, W., {Baliunas}, S., {et~al.} 2017, \apj, 835,
  \dodoi{10.3847/1538-4357/835/1/25}

\bibitem[{{Fabricius} {et~al.}(2002){Fabricius}, {H{\o}g}, {Makarov}, {Mason},
  {Wycoff}, \& {Urban}}]{Fabricius2002}
{Fabricius}, C., {H{\o}g}, E., {Makarov}, V.~V., {et~al.} 2002, \aap, 384, 180,
  \dodoi{10.1051/0004-6361:20011822}

\bibitem[{{Ferguson} {et~al.}(2005){Ferguson}, {Alexander}, {Allard}, {Barman},
  {Bodnarik}, {Hauschildt}, {Heffner-Wong}, \& {Tamanai}}]{Ferguson2005}
{Ferguson}, J.~W., {Alexander}, D.~R., {Allard}, F., {et~al.} 2005, \apj, 623,
  585, \dodoi{10.1086/428642}

\bibitem[{{Foreman-Mackey} {et~al.}(2013){Foreman-Mackey}, {Hogg}, {Lang}, \&
  {Goodman}}]{ForemanMackey2013}
{Foreman-Mackey}, D., {Hogg}, D.~W., {Lang}, D., \& {Goodman}, J. 2013, \pasp,
  125, 306, \dodoi{10.1086/670067}

\bibitem[{{Formicola} {et~al.}(2004){Formicola}, {Imbriani}, {Costantini},
  {Angulo}, {Bemmerer}, {Bonetti}, {Broggini}, {Corvisiero}, {Cruz},
  {Descouvemont}, {F{\"u}l{\"o}p}, {Gervino}, {Guglielmetti}, {Gustavino},
  {Gy{\"u}rky}, {Jesus}, {Junker}, {Lemut}, {Menegazzo}, {Prati}, {Roca},
  {Rolfs}, {Romano}, {Rossi Alvarez}, {Sch{\"u}mann}, {Somorjai}, {Straniero},
  {Strieder}, {Terrasi}, {Trautvetter}, {Vomiero}, \&
  {Zavatarelli}}]{Formicola2004}
{Formicola}, A., {Imbriani}, G., {Costantini}, H., {et~al.} 2004, Physics
  Letters B, 591, 61, \dodoi{10.1016/j.physletb.2004.03.092}

\bibitem[{{Fuhrmann}(2008)}]{Fuhrmann2008}
{Fuhrmann}, K. 2008, \mnras, 384, 173, \dodoi{10.1111/j.1365-2966.2007.12671.x}

\bibitem[{{Gaia Collaboration} {et~al.}(2018){Gaia Collaboration}, {Brown},
  {Vallenari}, {Prusti}, {de Bruijne}, {Babusiaux}, {Bailer-Jones}, {Biermann},
  {Evans}, {Eyer}, {Jansen}, {Jordi}, {Klioner}, {Lammers}, {Lindegren},
  {Luri}, {Mignard}, {Panem}, {Pourbaix}, {Randich}, {Sartoretti}, {Siddiqui},
  {Soubiran}, {van Leeuwen}, {Walton}, {Arenou}, {Bastian}, {Cropper},
  {Drimmel}, {Katz}, {Lattanzi}, {Bakker}, {Cacciari}, {Casta{\~n}eda},
  {Chaoul}, {Cheek}, {De Angeli}, {Fabricius}, {Guerra}, {Holl}, {Masana},
  {Messineo}, {Mowlavi}, {Nienartowicz}, {Panuzzo}, {Portell}, {Riello},
  {Seabroke}, {Tanga}, {Th{\'e}venin}, {Gracia-Abril}, {Comoretto},
  {Garcia-Reinaldos}, {Teyssier}, {Altmann}, {Andrae}, {Audard},
  {Bellas-Velidis}, {Benson}, {Berthier}, {Blomme}, {Burgess}, {Busso},
  {Carry}, {Cellino}, {Clementini}, {Clotet}, {Creevey}, {Davidson}, {De
  Ridder}, {Delchambre}, {Dell'Oro}, {Ducourant},
  {Fern{\'a}ndez-Hern{\'a}ndez}, {Fouesneau}, {Fr{\'e}mat}, {Galluccio},
  {Garc{\'\i}a-Torres}, {Gonz{\'a}lez-N{\'u}{\~n}ez}, {Gonz{\'a}lez-Vidal},
  {Gosset}, {Guy}, {Halbwachs}, {Hambly}, {Harrison}, {Hern{\'a}ndez},
  {Hestroffer}, {Hodgkin}, {Hutton}, {Jasniewicz}, {Jean-Antoine-Piccolo},
  {Jordan}, {Korn}, {Krone-Martins}, {Lanzafame}, {Lebzelter}, {L{\"o}ffler},
  {Manteiga}, {Marrese}, {Mart{\'\i}n-Fleitas}, {Moitinho}, {Mora}, {Muinonen},
  {Osinde}, {Pancino}, {Pauwels}, {Petit}, {Recio-Blanco}, {Richards},
  {Rimoldini}, {Robin}, {Sarro}, {Siopis}, {Smith}, {Sozzetti}, {S{\"u}veges},
  {Torra}, {van Reeven}, {Abbas}, {Abreu Aramburu}, {Accart}, {Aerts},
  {Altavilla}, {{\'A}lvarez}, {Alvarez}, {Alves}, {Anderson}, {Andrei},
  {Anglada Varela}, {Antiche}, {Antoja}, {Arcay}, {Astraatmadja}, {Bach},
  {Baker}, {Balaguer-N{\'u}{\~n}ez}, {Balm}, {Barache}, {Barata}, {Barbato},
  {Barblan}, {Barklem}, {Barrado}, {Barros}, {Barstow}, {Bartholom{\'e}
  Mu{\~n}oz}, {Bassilana}, {Becciani}, {Bellazzini}, {Berihuete}, {Bertone},
  {Bianchi}, {Bienaym{\'e}}, {Blanco-Cuaresma}, {Boch}, {Boeche}, {Bombrun},
  {Borrachero}, {Bossini}, {Bouquillon}, {Bourda}, {Bragaglia}, {Bramante},
  {Breddels}, {Bressan}, {Brouillet}, {Br{\"u}semeister}, {Brugaletta},
  {Bucciarelli}, {Burlacu}, {Busonero}, {Butkevich}, {Buzzi}, {Caffau},
  {Cancelliere}, {Cannizzaro}, {Cantat-Gaudin}, {Carballo}, {Carlucci},
  {Carrasco}, {Casamiquela}, {Castellani}, {Castro-Ginard}, {Charlot},
  {Chemin}, {Chiavassa}, {Cocozza}, {Costigan}, {Cowell}, {Crifo}, {Crosta},
  {Crowley}, {Cuypers}, {Dafonte}, {Damerdji}, {Dapergolas}, {David}, {David},
  {de Laverny}, {De Luise}, {De March}, {de Martino}, {de Souza}, {de Torres},
  {Debosscher}, {del Pozo}, {Delbo}, {Delgado}, {Delgado}, {Di Matteo},
  {Diakite}, {Diener}, {Distefano}, {Dolding}, {Drazinos}, {Dur{\'a}n},
  {Edvardsson}, {Enke}, {Eriksson}, {Esquej}, {Eynard Bontemps}, {Fabre},
  {Fabrizio}, {Faigler}, {Falc{\~a}o}, {Farr{\`a}s Casas}, {Federici},
  {Fedorets}, {Fernique}, {Figueras}, {Filippi}, {Findeisen}, {Fonti},
  {Fraile}, {Fraser}, {Fr{\'e}zouls}, {Gai}, {Galleti}, {Garabato},
  {Garc{\'\i}a-Sedano}, {Garofalo}, {Garralda}, {Gavel}, {Gavras}, {Gerssen},
  {Geyer}, {Giacobbe}, {Gilmore}, {Girona}, {Giuffrida}, {Glass}, {Gomes},
  {Granvik}, {Gueguen}, {Guerrier}, {Guiraud}, {Guti{\'e}rrez-S{\'a}nchez},
  {Haigron}, {Hatzidimitriou}, {Hauser}, {Haywood}, {Heiter}, {Helmi}, {Heu},
  {Hilger}, {Hobbs}, {Hofmann}, {Holland}, {Huckle}, {Hypki}, {Icardi},
  {Jan{\ss}en}, {Jevardat de Fombelle}, {Jonker}, {Juh{\'a}sz}, {Julbe},
  {Karampelas}, {Kewley}, {Klar}, {Kochoska}, {Kohley}, {Kolenberg},
  {Kontizas}, {Kontizas}, {Koposov}, {Kordopatis}, {Kostrzewa-Rutkowska},
  {Koubsky}, {Lambert}, {Lanza}, {Lasne}, {Lavigne}, {Le Fustec}, {Le
  Poncin-Lafitte}, {Lebreton}, {Leccia}, {Leclerc}, {Lecoeur-Taibi},
  {Lenhardt}, {Leroux}, {Liao}, {Licata}, {Lindstr{\o}m}, {Lister}, {Livanou},
  {Lobel}, {L{\'o}pez}, {Managau}, {Mann}, {Mantelet}, {Marchal}, {Marchant},
  {Marconi}, {Marinoni}, {Marschalk{\'o}}, {Marshall}, {Martino}, {Marton},
  {Mary}, {Massari}, {Matijevi{\v{c}}}, {Mazeh}, {McMillan}, {Messina},
  {Michalik}, {Millar}, {Molina}, {Molinaro}, {Moln{\'a}r}, {Montegriffo},
  {Mor}, {Morbidelli}, {Morel}, {Morris}, {Mulone}, {Muraveva}, {Musella},
  {Nelemans}, {Nicastro}, {Noval}, {O'Mullane}, {Ord{\'e}novic},
  {Ord{\'o}{\~n}ez-Blanco}, {Osborne}, {Pagani}, {Pagano}, {Pailler},
  {Palacin}, {Palaversa}, {Panahi}, {Pawlak}, {Piersimoni}, {Pineau}, {Plachy},
  {Plum}, {Poggio}, {Poujoulet}, {Pr{\v{s}}a}, {Pulone}, {Racero}, {Ragaini},
  {Rambaux}, {Ramos-Lerate}, {Regibo}, {Reyl{\'e}}, {Riclet}, {Ripepi}, {Riva},
  {Rivard}, {Rixon}, {Roegiers}, {Roelens}, {Romero-G{\'o}mez}, {Rowell},
  {Royer}, {Ruiz-Dern}, {Sadowski}, {Sagrist{\`a} Sell{\'e}s}, {Sahlmann},
  {Salgado}, {Salguero}, {Sanna}, {Santana-Ros}, {Sarasso}, {Savietto},
  {Schultheis}, {Sciacca}, {Segol}, {Segovia}, {S{\'e}gransan}, {Shih},
  {Siltala}, {Silva}, {Smart}, {Smith}, {Solano}, {Solitro}, {Sordo}, {Soria
  Nieto}, {Souchay}, {Spagna}, {Spoto}, {Stampa}, {Steele},
  {Steidelm{\"u}ller}, {Stephenson}, {Stoev}, {Suess}, {Surdej}, {Szabados},
  {Szegedi-Elek}, {Tapiador}, {Taris}, {Tauran}, {Taylor}, {Teixeira},
  {Terrett}, {Teyssand ier}, {Thuillot}, {Titarenko}, {Torra Clotet}, {Turon},
  {Ulla}, {Utrilla}, {Uzzi}, {Vaillant}, {Valentini}, {Valette}, {van Elteren},
  {Van Hemelryck}, {van Leeuwen}, {Vaschetto}, {Vecchiato}, {Veljanoski},
  {Viala}, {Vicente}, {Vogt}, {von Essen}, {Voss}, {Votruba}, {Voutsinas},
  {Walmsley}, {Weiler}, {Wertz}, {Wevers}, {Wyrzykowski}, {Yoldas},
  {{\v{Z}}erjal}, {Ziaeepour}, {Zorec}, {Zschocke}, {Zucker}, {Zurbach}, \&
  {Zwitter}}]{Gaia2018}
{Gaia Collaboration}, {Brown}, A.~G.~A., {Vallenari}, A., {et~al.} 2018, \aap,
  616, A1, \dodoi{10.1051/0004-6361/201833051}

\bibitem[{{Garc{\'\i}a} {et~al.}(2014){Garc{\'\i}a}, {Ceillier}, {Salabert},
  {Mathur}, {van Saders}, {Pinsonneault}, {Ballot}, {Beck}, {Bloemen},
  {Campante}, {Davies}, {do Nascimento}, {Mathis}, {Metcalfe}, {Nielsen},
  {Su{\'a}rez}, {Chaplin}, {Jim{\'e}nez}, \& {Karoff}}]{Garcia2014}
{Garc{\'\i}a}, R.~A., {Ceillier}, T., {Salabert}, D., {et~al.} 2014, \aap, 572,
  A34, \dodoi{10.1051/0004-6361/201423888}

\bibitem[{{Gray} {et~al.}(2006){Gray}, {Corbally}, {Garrison}, {McFadden},
  {Bubar}, {McGahee}, {O'Donoghue}, \& {Knox}}]{Gray2006}
{Gray}, R.~O., {Corbally}, C.~J., {Garrison}, R.~F., {et~al.} 2006, \aj, 132,
  161, \dodoi{10.1086/504637}

\bibitem[{{Grevesse} \& {Sauval}(1998)}]{Grevesse1998}
{Grevesse}, N., \& {Sauval}, A.~J. 1998, \ssr, 85, 161,
  \dodoi{10.1023/A:1005161325181}

\bibitem[{{Hall} {et~al.}(2007){Hall}, {Lockwood}, \& {Skiff}}]{Hall2007}
{Hall}, J.~C., {Lockwood}, G.~W., \& {Skiff}, B.~A. 2007, \aj, 133, 862,
  \dodoi{10.1086/510356}

\bibitem[{{Harvey} {et~al.}(1988){Harvey}, {Hill}, {Kennedy}, {Leibacher}, \&
  {Livingston}}]{Harvey1988}
{Harvey}, J.~W., {Hill}, F., {Kennedy}, J.~R., {Leibacher}, J.~W., \&
  {Livingston}, W.~C. 1988, Advances in Space Research, 8, 117,
  \dodoi{10.1016/0273-1177(88)90304-3}

\bibitem[{{Iglesias} \& {Rogers}(1996)}]{Iglesias1996}
{Iglesias}, C.~A., \& {Rogers}, F.~J. 1996, \apj, 464, 943,
  \dodoi{10.1086/177381}

\bibitem[{{Jenkins} {et~al.}(2016){Jenkins}, {Twicken}, {McCauliff},
  {Campbell}, {Sanderfer}, {Lung}, {Mansouri-Samani}, {Girouard}, {Tenenbaum},
  {Klaus}, {Smith}, {Caldwell}, {Chacon}, {Henze}, {Heiges}, {Latham},
  {Morgan}, {Swade}, {Rinehart}, \& {Vanderspek}}]{Jenkins2016}
{Jenkins}, J.~M., {Twicken}, J.~D., {McCauliff}, S., {et~al.} 2016, in
  \procspie, Vol. 9913, Software and Cyberinfrastructure for Astronomy IV,
  99133E

\bibitem[{{Judge} {et~al.}(2004){Judge}, {Saar}, {Carlsson}, \&
  {Ayres}}]{Judge2004}
{Judge}, P.~G., {Saar}, S.~H., {Carlsson}, M., \& {Ayres}, T.~R. 2004, \apj,
  609, 392, \dodoi{10.1086/421044}

\bibitem[{{Katoh} {et~al.}(2013){Katoh}, {Itoh}, {Toyota}, \&
  {Sato}}]{Katoh2013}
{Katoh}, N., {Itoh}, Y., {Toyota}, E., \& {Sato}, B. 2013, \aj, 145, 41,
  \dodoi{10.1088/0004-6256/145/2/41}

\bibitem[{{Kraft}(1967)}]{Kraft1967}
{Kraft}, R.~P. 1967, \apj, 150, 551, \dodoi{10.1086/149359}

\bibitem[{{Kurucz}(1992)}]{Kurucz1992}
{Kurucz}, R.~L. 1992, in IAU Symposium, Vol. 149, The Stellar Populations of
  Galaxies, ed. B.~{Barbuy} \& A.~{Renzini}, 225

\bibitem[{{Li} {et~al.}(2019){Li}, {Bedding}, {Kjeldsen}, {Stello},
  {Christensen-Dalsgaard}, \& {Deng}}]{Li2019}
{Li}, T., {Bedding}, T.~R., {Kjeldsen}, H., {et~al.} 2019, \mnras, 483, 780,
  \dodoi{10.1093/mnras/sty3000}

\bibitem[{{Linsky} \& {Avrett}(1970)}]{Linsky1970}
{Linsky}, J.~L., \& {Avrett}, E.~H. 1970, \pasp, 82, 169,
  \dodoi{10.1086/128904}

\bibitem[{{Lund} {et~al.}(2015){Lund}, {Handberg}, {Davies}, {Chaplin}, \&
  {Jones}}]{Lund2015}
{Lund}, M.~N., {Handberg}, R., {Davies}, G.~R., {Chaplin}, W.~J., \& {Jones},
  C.~D. 2015, \apj, 806, 30, \dodoi{10.1088/0004-637X/806/1/30}

\bibitem[{{Lund} {et~al.}(2017){Lund}, {Handberg}, {Kjeldsen}, {Chaplin}, \&
  {Christensen-Dalsgaard}}]{Lund2017}
{Lund}, M.~N., {Handberg}, R., {Kjeldsen}, H., {Chaplin}, W.~J., \&
  {Christensen-Dalsgaard}, J. 2017, in European Physical Journal Web of
  Conferences, Vol. 160, European Physical Journal Web of Conferences, 01005

\bibitem[{{MacGregor} \& {Brenner}(1991)}]{MacGregor1991}
{MacGregor}, K.~B., \& {Brenner}, M. 1991, \apj, 376, 204,
  \dodoi{10.1086/170269}

\bibitem[{{Magic} {et~al.}(2015){Magic}, {Weiss}, \& {Asplund}}]{magic2015}
{Magic}, Z., {Weiss}, A., \& {Asplund}, M. 2015, \aap, 573, A89,
  \dodoi{10.1051/0004-6361/201423760}

\bibitem[{{Marcadon} {et~al.}(2018){Marcadon}, {Appourchaux}, \&
  {Marques}}]{Marcadon2018}
{Marcadon}, F., {Appourchaux}, T., \& {Marques}, J.~P. 2018, \aap, 617, A2,
  \dodoi{10.1051/0004-6361/201731628}

\bibitem[{{Mason} {et~al.}(2001){Mason}, {Wycoff}, {Hartkopf}, {Douglass}, \&
  {Worley}}]{Mason2001}
{Mason}, B.~D., {Wycoff}, G.~L., {Hartkopf}, W.~I., {Douglass}, G.~G., \&
  {Worley}, C.~E. 2001, \aj, 122, 3466, \dodoi{10.1086/323920}

\bibitem[{{Mathur} {et~al.}(2010){Mathur}, {Garc{\'\i}a}, {R{\'e}gulo},
  {Creevey}, {Ballot}, {Salabert}, {Arentoft}, {Quirion}, {Chaplin}, \&
  {Kjeldsen}}]{Mathur2010}
{Mathur}, S., {Garc{\'\i}a}, R.~A., {R{\'e}gulo}, C., {et~al.} 2010, \aap, 511,
  A46, \dodoi{10.1051/0004-6361/200913266}

\bibitem[{{Mermilliod}(2006)}]{Mermilliod2006}
{Mermilliod}, J.~C. 2006, VizieR Online Data Catalog, II/168

\bibitem[{{Metcalfe} {et~al.}(2019){Metcalfe}, {Kochukhov}, {Ilyin},
  {Strassmeier}, {Godoy-Rivera}, \& {Pinsonneault}}]{Metcalfe2019b}
{Metcalfe}, T.~S., {Kochukhov}, O., {Ilyin}, I.~V., {et~al.} 2019, \apjl, 887,
  L38, \dodoi{10.3847/2041-8213/ab5e48}

\bibitem[{{Metcalfe} \& {van Saders}(2017)}]{Metcalfe2017}
{Metcalfe}, T.~S., \& {van Saders}, J. 2017, \solphys, 292, 126,
  \dodoi{10.1007/s11207-017-1157-5}

\bibitem[{{Montgomery} \& {O'Donoghue}(1999)}]{Montgomery1999}
{Montgomery}, M.~H., \& {O'Donoghue}, D. 1999, Delta Scuti Star Newsletter, 13,
  28

\bibitem[{{Mosser} {et~al.}(2015){Mosser}, {Vrard}, {Belkacem}, {Deheuvels}, \&
  {Goupil}}]{Mosser2015}
{Mosser}, B., {Vrard}, M., {Belkacem}, K., {Deheuvels}, S., \& {Goupil}, M.~J.
  2015, \aap, 584, A50, \dodoi{10.1051/0004-6361/201527075}

\bibitem[{{Mosser} {et~al.}(2014){Mosser}, {Benomar}, {Belkacem}, {Goupil},
  {Lagarde}, {Michel}, {Lebreton}, {Stello}, {Vrard}, {Barban}, {Bedding},
  {Deheuvels}, {Chaplin}, {De Ridder}, {Elsworth}, {Montalban}, {Noels},
  {Ouazzani}, {Samadi}, {White}, \& {Kjeldsen}}]{Mosser2014}
{Mosser}, B., {Benomar}, O., {Belkacem}, K., {et~al.} 2014, \aap, 572, L5,
  \dodoi{10.1051/0004-6361/201425039}

\bibitem[{{Olspert} {et~al.}(2018){Olspert}, {Lehtinen}, {K{\"a}pyl{\"a}},
  {Pelt}, \& {Grigorievskiy}}]{Olspert2018}
{Olspert}, N., {Lehtinen}, J.~J., {K{\"a}pyl{\"a}}, M.~J., {Pelt}, J., \&
  {Grigorievskiy}, A. 2018, \aap, 619, A6, \dodoi{10.1051/0004-6361/201732525}

\bibitem[{{Paunzen}(2015)}]{Paunzen2015}
{Paunzen}, E. 2015, \aap, 580, A23, \dodoi{10.1051/0004-6361/201526413}

\bibitem[{{Paxton} {et~al.}(2011){Paxton}, {Bildsten}, {Dotter}, {Herwig},
  {Lesaffre}, \& {Timmes}}]{Paxton2011}
{Paxton}, B., {Bildsten}, L., {Dotter}, A., {et~al.} 2011, \apjs, 192, 3,
  \dodoi{10.1088/0067-0049/192/1/3}

\bibitem[{{Ricker} {et~al.}(2014){Ricker}, {Winn}, {Vanderspek}, {Latham},
  {Bakos}, {Bean}, {Berta-Thompson}, {Brown}, {Buchhave}, {Butler}, {Butler},
  {Chaplin}, {Charbonneau}, {Christensen-Dalsgaard}, {Clampin}, {Deming},
  {Doty}, {De Lee}, {Dressing}, {Dunham}, {Endl}, {Fressin}, {Ge}, {Henning},
  {Holman}, {Howard}, {Ida}, {Jenkins}, {Jernigan}, {Johnson}, {Kaltenegger},
  {Kawai}, {Kjeldsen}, {Laughlin}, {Levine}, {Lin}, {Lissauer}, {MacQueen},
  {Marcy}, {McCullough}, {Morton}, {Narita}, {Paegert}, {Palle}, {Pepe},
  {Pepper}, {Quirrenbach}, {Rinehart}, {Sasselov}, {Sato}, {Seager},
  {Sozzetti}, {Stassun}, {Sullivan}, {Szentgyorgyi}, {Torres}, {Udry}, \&
  {Villasenor}}]{Ricker2014}
{Ricker}, G.~R., {Winn}, J.~N., {Vanderspek}, R., {et~al.} 2014, in Society of
  Photo-Optical Instrumentation Engineers (SPIE) Conference Series, Vol. 9143,
  Proceedings of the SPIE, Volume 9143, id. 914320 15 pp. (2014)., 914320

\bibitem[{{Rogers} \& {Nayfonov}(2002)}]{Rogers2002}
{Rogers}, F.~J., \& {Nayfonov}, A. 2002, \apj, 576, 1064,
  \dodoi{10.1086/341894}

\bibitem[{{Saio} {et~al.}(2015){Saio}, {Kurtz}, {Takata}, {Shibahashi},
  {Murphy}, {Sekii}, \& {Bedding}}]{Saio2015}
{Saio}, H., {Kurtz}, D.~W., {Takata}, M., {et~al.} 2015, \mnras, 447, 3264,
  \dodoi{10.1093/mnras/stu2696}

\bibitem[{{Sarma}(1962)}]{Sarma1962}
{Sarma}, M.~B.~K. 1962, \apj, 135, 301, \dodoi{10.1086/147268}

\bibitem[{{Schofield} {et~al.}(2019){Schofield}, {Chaplin}, {Huber},
  {Campante}, {Davies}, {Miglio}, {Ball}, {Appourchaux}, {Basu}, {Bedding},
  {Christensen-Dalsgaard}, {Creevey}, {Garc{\'\i}a}, {Handberg}, {Kawaler},
  {Kjeldsen}, {Latham}, {Lund}, {Metcalfe}, {Ricker}, {Serenelli}, {Silva
  Aguirre}, {Stello}, \& {Vanderspek}}]{Schofield2019}
{Schofield}, M., {Chaplin}, W.~J., {Huber}, D., {et~al.} 2019, \apjs, 241, 12,
  \dodoi{10.3847/1538-4365/ab04f5}

\bibitem[{{Serenelli} {et~al.}(2017){Serenelli}, {Johnson}, {Huber},
  {Pinsonneault}, {Ball}, {Tayar}, {Silva Aguirre}, {Basu}, {Troup}, {Hekker},
  {Kallinger}, {Stello}, {Davies}, {Lund}, {Mathur}, {Mosser}, {Stassun},
  {Chaplin}, {Elsworth}, {Garc{\'\i}a}, {Handberg}, {Holtzman}, {Hearty},
  {Garc{\'\i}a-Hern{\'a}ndez}, {Gaulme}, \& {Zamora}}]{Serenelli2017}
{Serenelli}, A., {Johnson}, J., {Huber}, D., {et~al.} 2017, \apjs, 233, 23,
  \dodoi{10.3847/1538-4365/aa97df}

\bibitem[{{Shibahashi}(1979)}]{Shibahashi1979}
{Shibahashi}, H. 1979, \pasj, 31, 87

\bibitem[{{Somers} \& {Pinsonneault}(2016)}]{Somers2016}
{Somers}, G., \& {Pinsonneault}, M.~H. 2016, \apj, 829, 32,
  \dodoi{10.3847/0004-637X/829/1/32}

\bibitem[{{Soon} {et~al.}(1994){Soon}, {Baliunas}, \& {Zhang}}]{Soon1994}
{Soon}, W.~H., {Baliunas}, S.~L., \& {Zhang}, Q. 1994, \solphys, 154, 385,
  \dodoi{10.1007/BF00681107}

\bibitem[{{Stassun} {et~al.}(2017){Stassun}, {Collins}, \&
  {Gaudi}}]{Stassun2017}
{Stassun}, K.~G., {Collins}, K.~A., \& {Gaudi}, B.~S. 2017, \aj, 153, 136,
  \dodoi{10.3847/1538-3881/aa5df3}

\bibitem[{{Stassun} {et~al.}(2018){Stassun}, {Corsaro}, {Pepper}, \&
  {Gaudi}}]{Stassun2018}
{Stassun}, K.~G., {Corsaro}, E., {Pepper}, J.~A., \& {Gaudi}, B.~S. 2018, \aj,
  155, 22, \dodoi{10.3847/1538-3881/aa998a}

\bibitem[{{Stassun} \& {Torres}(2016)}]{Stassun2016}
{Stassun}, K.~G., \& {Torres}, G. 2016, \apjl, 831, L6,
  \dodoi{10.3847/2041-8205/831/1/L6}

\bibitem[{{Stassun} \& {Torres}(2018)}]{StassunTorres2018}
---. 2018, \apj, 862, 61, \dodoi{10.3847/1538-4357/aacafc}

\bibitem[{{Steigman}(2010)}]{Steigman2010}
{Steigman}, G. 2010, \jcap, 4, 029.
\newblock \doarXiv{1002.3604}

\bibitem[{{Tayar} {et~al.}(2017){Tayar}, {Somers}, {Pinsonneault}, {Stello},
  {Mints}, {Johnson}, {Zamora}, {Garc{\'\i}a-Hern{\'a}ndez}, {Maraston},
  {Serenelli}, {Allende Prieto}, {Bastien}, {Basu}, {Bird}, {Cohen}, {Cunha},
  {Elsworth}, {Garc{\'\i}a}, {Girardi}, {Hekker}, {Holtzman}, {Huber},
  {Mathur}, {M{\'e}sz{\'a}ros}, {Mosser}, {Shetrone}, {Silva Aguirre},
  {Stassun}, {Stringfellow}, {Zasowski}, \& {Roman-Lopes}}]{tayar2017}
{Tayar}, J., {Somers}, G., {Pinsonneault}, M.~H., {et~al.} 2017, \apj, 840, 17,
  \dodoi{10.3847/1538-4357/aa6a1e}

\bibitem[{{Thompson} {et~al.}(1996){Thompson}, {Toomre}, {Anderson}, {Antia},
  {Berthomieu}, {Burtonclay}, {Chitre}, {Christensen-Dalsgaard}, {Corbard}, {De
  Rosa}, {Genovese}, {Gough}, {Haber}, {Harvey}, {Hill}, {Howe}, {Korzennik},
  {Kosovichev}, {Leibacher}, {Pijpers}, {Provost}, {Rhodes}, {Schou}, {Sekii},
  {Stark}, \& {Wilson}}]{Thompson1996}
{Thompson}, M.~J., {Toomre}, J., {Anderson}, E.~R., {et~al.} 1996, Science,
  272, 1300, \dodoi{10.1126/science.272.5266.1300}

\bibitem[{{Thoul} {et~al.}(1994){Thoul}, {Bahcall}, \& {Loeb}}]{Thoul1994}
{Thoul}, A.~A., {Bahcall}, J.~N., \& {Loeb}, A. 1994, \apj, 421, 828,
  \dodoi{10.1086/173695}

\bibitem[{{Tokovinin} {et~al.}(2015){Tokovinin}, {Mason}, {Hartkopf}, {Mendez},
  \& {Horch}}]{Tokovinin2015}
{Tokovinin}, A., {Mason}, B.~D., {Hartkopf}, W.~I., {Mendez}, R.~A., \&
  {Horch}, E.~P. 2015, \aj, 150, 50, \dodoi{10.1088/0004-6256/150/2/50}

\bibitem[{{Tripathi} {et~al.}(2018){Tripathi}, {Nandy}, \&
  {Banerjee}}]{Tripathi2018}
{Tripathi}, B., {Nandy}, D., \& {Banerjee}, S. 2018, arXiv e-prints,
  arXiv:1812.05533.
\newblock \doarXiv{1812.05533}

\bibitem[{{van Saders} {et~al.}(2016){van Saders}, {Ceillier}, {Metcalfe},
  {Silva Aguirre}, {Pinsonneault}, {Garc{\'\i}a}, {Mathur}, \&
  {Davies}}]{vanSaders2016}
{van Saders}, J.~L., {Ceillier}, T., {Metcalfe}, T.~S., {et~al.} 2016, \nat,
  529, 181, \dodoi{10.1038/nature16168}

\bibitem[{{van Saders} \& {Pinsonneault}(2013)}]{vanSaders2013}
{van Saders}, J.~L., \& {Pinsonneault}, M.~H. 2013, \apj, 776, 67,
  \dodoi{10.1088/0004-637X/776/2/67}

\bibitem[{{van Saders} {et~al.}(2019){van Saders}, {Pinsonneault}, \&
  {Barbieri}}]{vanSaders2019}
{van Saders}, J.~L., {Pinsonneault}, M.~H., \& {Barbieri}, M. 2019, \apj, 872,
  128, \dodoi{10.3847/1538-4357/aafafe}

\bibitem[{{VandenBerg} \& {Clem}(2003)}]{VandenBerg2003}
{VandenBerg}, D.~A., \& {Clem}, J.~L. 2003, \aj, 126, 778,
  \dodoi{10.1086/376840}

\bibitem[{{Vaughan} {et~al.}(1978){Vaughan}, {Preston}, \&
  {Wilson}}]{Vaughan1978}
{Vaughan}, A.~H., {Preston}, G.~W., \& {Wilson}, O.~C. 1978, \pasp, 90, 267,
  \dodoi{10.1086/130324}

\bibitem[{{Viani} {et~al.}(2018){Viani}, {Basu}, {Ong J.}, {Bonaca}, \&
  {Chaplin}}]{viani2018}
{Viani}, L.~S., {Basu}, S., {Ong J.}, M.~J., {Bonaca}, A., \& {Chaplin}, W.~J.
  2018, \apj, 858, 28, \dodoi{10.3847/1538-4357/aab7eb}

\bibitem[{{White} \& {Livingston}(1981)}]{White1981}
{White}, O.~R., \& {Livingston}, W.~C. 1981, \apj, 249, 798,
  \dodoi{10.1086/159338}

\bibitem[{{White} {et~al.}(2018){White}, {Huber}, {Mann}, {Casagrande},
  {Grunblatt}, {Justesen}, {Silva Aguirre}, {Bedding}, {Ireland}, {Schaefer},
  \& {Tuthill}}]{White2018}
{White}, T.~R., {Huber}, D., {Mann}, A.~W., {et~al.} 2018, \mnras, 477, 4403,
  \dodoi{10.1093/mnras/sty898}

\bibitem[{{Wilson}(1978)}]{Wilson1978}
{Wilson}, O.~C. 1978, \apj, 226, 379, \dodoi{10.1086/156618}

\bibitem[{{Zinn}(2019)}]{Zinn2018}
{Zinn}, J.~C. 2019, in American Astronomical Society Meeting Abstracts, Vol.
  233, American Astronomical Society Meeting Abstracts \#233, 341.07

\end{thebibliography}
\end{document}